\documentclass[conference,compsoc]{IEEEtran}
\ifCLASSOPTIONcompsoc
  \usepackage[nocompress]{cite}
\else
  \usepackage{cite}
\fi

\ifCLASSINFOpdf
   \usepackage[pdftex]{graphicx}
   \graphicspath{{images/}}
\else
  \usepackage[dvips]{graphicx}
  \graphicspath{{images/}}
\fi
\usepackage[cmex10]{amsmath}
\usepackage{listings}
\usepackage{color}

\usepackage{microtype}
\usepackage{graphicx}
\usepackage{subcaption}
\usepackage{booktabs} 

\usepackage{fvextra}
\usepackage{algorithmic}
\usepackage{hyperref}
\usepackage{booktabs}
\usepackage{multirow}
\usepackage[table]{xcolor}
\usepackage{hyperref}
\usepackage{amsfonts}
\usepackage{amsmath,amssymb}
\usepackage[dvipsnames]{xcolor} 
\newcommand{\best}[1]{\textcolor{BrickRed}{\mathbf{#1}}}
\newcommand{\secondbest}[1]{\textcolor{NavyBlue}{\underline{#1}}}
\newcommand{\thirdbest}[1]{\textcolor{ForestGreen}{#1}}
\usepackage{bm}

\definecolor{codegreen}{rgb}{0,0.6,0}
\definecolor{codegray}{rgb}{0.5,0.5,0.5}
\definecolor{codepurple}{rgb}{0.58,0,0.82}
\definecolor{backcolour}{rgb}{0.95,0.95,0.92}
 
\lstdefinestyle{mystyle}{
    backgroundcolor=\color{backcolour},   
    commentstyle=\color{codegreen},
    keywordstyle=\color{magenta},
    numberstyle=\tiny\color{codegray},
    stringstyle=\color{codepurple},
    basicstyle=\footnotesize,
    breakatwhitespace=false,         
    breaklines=true,                 
    captionpos=b,                    
    keepspaces=true,                 
    numbers=left,                    
    numbersep=5pt,                  
    showspaces=false,                
    showstringspaces=false,
    showtabs=false,
    tabsize=2
}
 
\begin{document}
\title{MotionDuet: Dual-Conditioned 3D Human Motion Generation with Video-Regularized Text Learning}

\author{
\IEEEauthorblockN{
YiYang Zhang\textsuperscript{1, *},
Tengjiao Sun\textsuperscript{2, *},
Pengcheng Fang\textsuperscript{2, *},
Deng-Bao Wang\textsuperscript{1}, \\
Xiaohao Cai\textsuperscript{2},
Min-Ling Zhang\textsuperscript{1},
Hansung Kim\textsuperscript{2}
}

\IEEEauthorblockA{
\textsuperscript{1}Southeast University \quad
\textsuperscript{2}University of Southampton
}


\IEEEauthorblockA{
\textsuperscript{*}Equal contribution. \quad
\textsuperscript{\textdagger}Corresponding author.
}
}

\twocolumn[{%
\renewcommand\twocolumn[1][]{#1}%
\maketitle
\begin{center}
  \includegraphics[width=\textwidth]{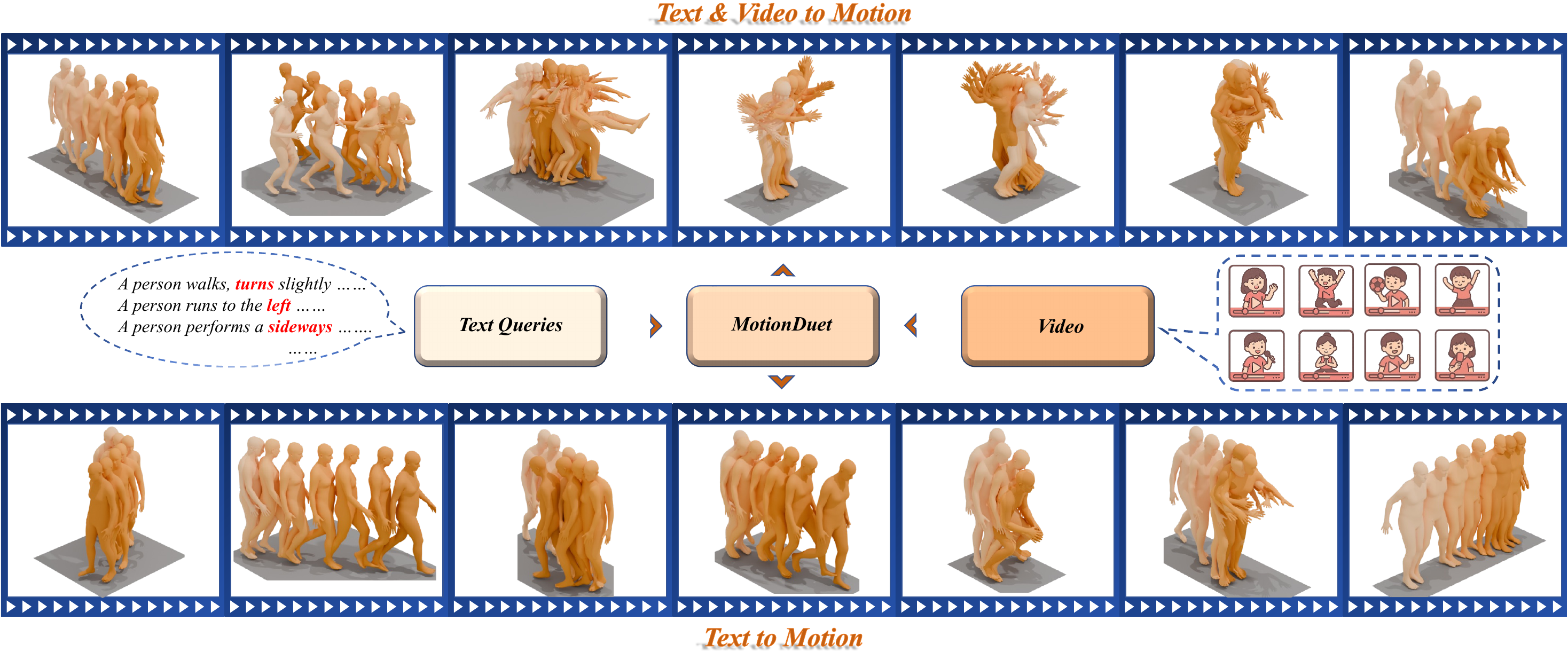}
  \captionof{figure}{MotionDuet is a multimodal framework for generating high-quality, controllable human motion under diverse conditions, including text prompts, video references, or their combination. Video results are provided in the supplementary material.}
  \label{fig:figure1}
\end{center}
}]


\begin{abstract}
3D Human motion generation is pivotal across film, animation, gaming, and embodied intelligence. 
Traditional 3D motion synthesis relies on costly motion capture, while recent work shows that 2D videos provide rich, temporally coherent observations of human behavior. 
Existing approaches, however, either map high-level text descriptions to motion or rely solely on video conditioning, leaving a gap between generated dynamics and real-world motion statistics. 
We introduce \textbf{MotionDuet}, a multimodal framework that aligns motion generation with the distribution of video-derived representations. 
In this dual-conditioning paradigm, video cues extracted from a pretrained model (e.g., VideoMAE) ground low-level motion dynamics, while textual prompts provide semantic intent. 
To bridge the distribution gap across modalities, we propose \emph{Dual-stream Unified Encoding and Transformation} (\textbf{DUET}) and a \emph{Distribution-Aware Structural Harmonization} (\textbf{DASH}) loss. 
DUET fuses video-informed cues into the motion latent space via unified encoding and dynamic attention, while DASH aligns motion trajectories with both distributional and structural statistics of video features. 
An \emph{auto-guidance} mechanism further balances textual and visual signals by leveraging a weakened copy of the model, enhancing controllability without sacrificing diversity. 
Extensive experiments demonstrate that MotionDuet generates realistic and controllable human motions, surpassing strong state-of-the-art baselines.

\end{abstract}

\section{Introduction}
Generating high-quality 3D  human motion from textual or visual inputs is a central challenge in vision, graphics, and embodied AI \cite{zhang2024motiondiffuse,mou2024revideo}. This task underpins a broad range of applications such as virtual character animation, interactive systems, and robot teleoperation. Text-conditioned models excel at capturing semantic intent but often struggle to produce temporally coherent and physically plausible motion sequences \cite{baumann2025continuous,wang2025fg}. In contrast, video-conditioned models can accurately reproduce observed trajectories \cite{jeong2024vmc,mou2024revideo}, yet they require videos at inference time and tend to generalize poorly beyond training distributions. 

Both motion estimation and motion generation hinge on modeling human dynamics and temporal coherence \cite{liang2024intergen}. Recent advances in cross-task transfer suggest that distributional priors learned from robust representations such as DINOv2 \cite{oquab2023dinov2} can regularize generative models and improve physical consistency \cite{yu2024representation}. Inspired by this insight, we unify textual semantics and video cues within a coherent multimodal framework in which real-world video statistics inform the latent representation of motion. By aligning the distribution of motion embeddings with the distribution of video features extracted from a pretrained foundation model such as VideoMAE, our method enables the generator to inherit the natural variability of real human dynamics while staying faithful to textual intent.

In this work, we present \textbf{MotionDuet}, a multimodal 3D human motion generation paradigm inspired by theatrical direction. MotionDuet fuses video and text cues through a dual-conditioning scheme: the video branch, derived from VideoMAE embeddings, grounds motion trajectory and style, while the text branch conveys high-level intent. Importantly, the dual-modal training not only enables controllable generation when both inputs are available, but also significantly enhances the model’s ability to synthesize realistic and coherent motions from text alone. This demonstrates that video-conditioned supervision serves as an effective regularizer, transferring spatio-temporal priors from real videos to improve text-conditioned motion generation. Examples of multimodal inputs and generated motions are shown in Fig.~\ref{fig:figure1}.

To enable MotionDuet to learn realistic, semantically consistent, and controllable motion generation, we introduce three core designs. The main contributions of this paper are summarized as follows:
\begin{itemize}
\item First, the \emph{Distribution-Aware Structural Harmonization} (DASH) loss bridges the distributional gap between video representations and motion embeddings by aligning the motion latent space with real video features through token-level and structural consistency regularization. 

\item Second, the \emph{Dual-stream Unified Encoding and Transformation} (DUET) module integrates motion, textual, and visual cues through dynamic attention, frequency-domain reasoning, and similarity-based selection, thereby enhancing multimodal interaction and controllability. 

\item Third, an \emph{auto-guidance} strategy employs a degraded model copy to stabilize training and balance text–video conditioning signals. 
Although MotionDuet is trained under multimodal supervision, it does not rely on video input at inference. Instead, the inclusion of video-conditioned learning serves as a powerful regularizer that transfers real-world spatio-temporal priors into the motion latent space, substantially improving realism and coherence even when only textual input is provided. 
\end{itemize}
Through these designs, MotionDuet learns a robust and generalized motion prior that captures the intrinsic dynamics of human movement rather than merely replicating observed visual cues, enabling flexible inference under both text-only and multimodal conditions.

\section{Related Work}
\label{sec:relax}
\vspace{3pt}\noindent\textbf{Human Motion Generation.}
Human motion generation utilizes multimodal inputs such as text~\cite{zou2024parco,sheng2024exploring}, images~\cite{chen2022learning}, and music~\cite{wang2024dancecamera3d,zhang2024bidirectional}. Common tasks include unconditional motion generation~\cite{Raab_2023_CVPR} and text-conditioned generation~\cite{Wang_2023_ICCV}, where sequence-to-sequence models like Hier \cite{ghosh2021synthesis} improve realism.  Diffusion models further enhance sample quality and diversity, with MotionDiffuse~\cite{zhang2024motiondiffuse} enabling diverse synthesis via probabilistic modeling. GPT-based models, exemplified by MotionGPT \cite{jiang2024motiongpt}, discretize 3D motions into tokens and integrate them with text to improve performance across motion tasks. Mask-based frameworks also made significant strides last year; for example,  MoMask and Mogo \cite{guo2024momask, fu2026mogo} introduces hierarchical discrete representations and two-stage modeling.

\vspace{3pt}\noindent\textbf{Representation Learning.}
Motion representations are typically based on either SMPL parameters or hand-crafted features. The SMPL-based approach models motion by manipulating pose and shape parameters to generate 3D human meshes~\cite{cai2024smpler,loper2023smpl,wang2024disentangled,cao2023sesdf}. Alternatively, hand-crafted features~\cite{guo2022generating,starke2022deepphase,chen2023executing} are designed to address animation artifacts like foot sliding, improving realism and control in motion synthesis.

\vspace{3pt}\noindent\textbf{Multimodal Condition.}
Adapters, controllers, and classifier-free guidance (CFG) are widely used to enhance multimodal generative models. Adapters such as MCRE~\cite{sun2024mcre} enable efficient modality adaptation (e.g., text-to-motion) via lightweight modules in CLIP space. Controllers improve controllability without additional parameters, as demonstrated in TLControl~\cite{wan2024tlcontrol}. CFG~\cite{shen2024rethinking,kwon2025tcfg} guides diffusion models toward high-quality conditional generation, especially in text-to-image tasks. Together, these mechanisms significantly improve flexibility and generation quality in multimodal settings.

\begin{figure*}[htbp]
    \centering
    \includegraphics[width=0.93\linewidth]{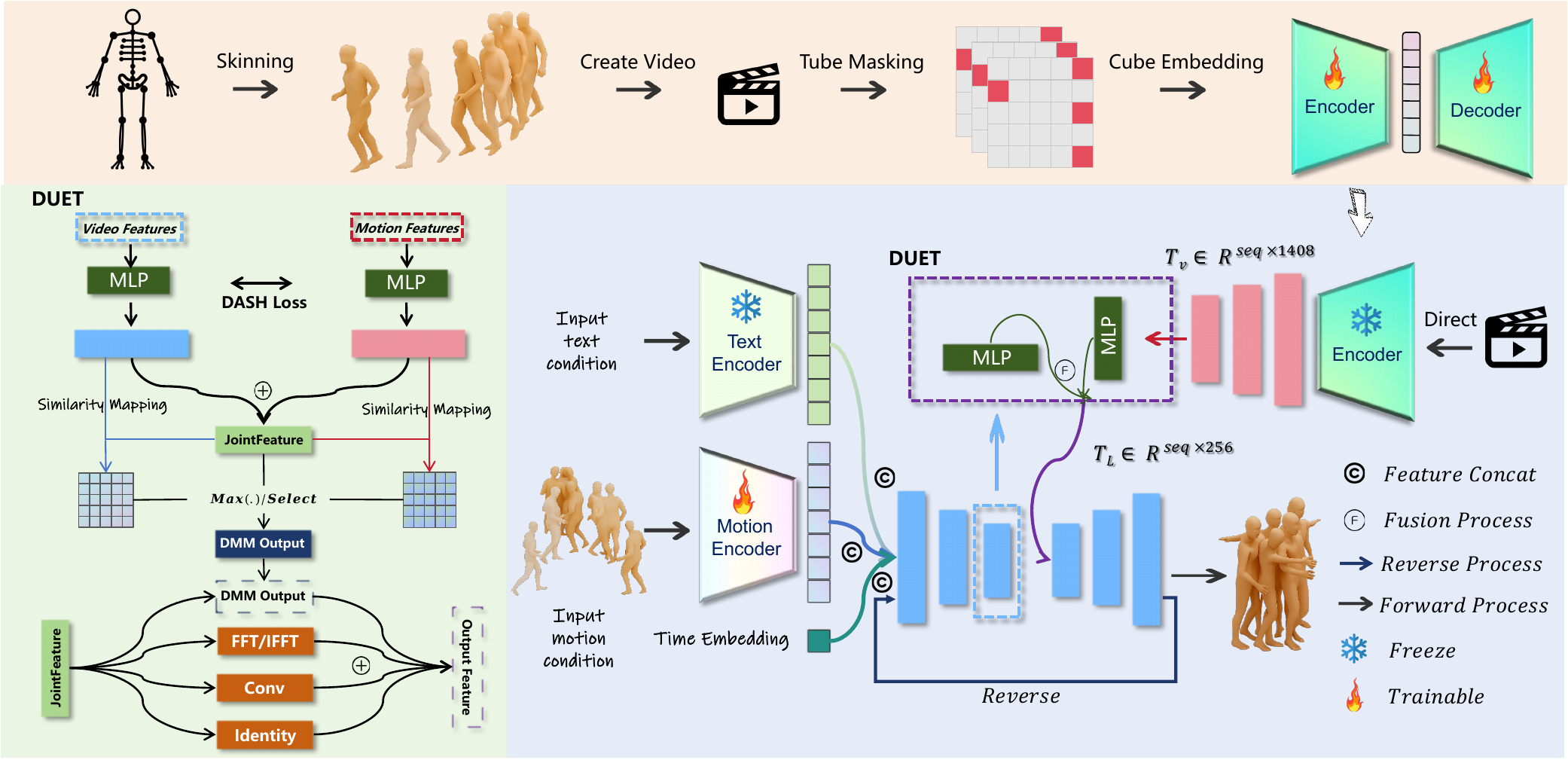}
    \caption{MotionDuet framework overview. It primarily consists of three key steps: 1) fine-tuning video motion dataset based on a pre-trained model and freezing the weights to focus on inference (orange background); 2) proposing a dual-stream control mechanism combined with auto-guidance mechanism to integrate video and text inputs, effectively guiding motion generation (blue background); and 3) utilizing the DUET module (purple dashed box) combined with DASH Loss to align and fuse multimodal information, enhancing overall information processing capabilities.}
    \label{fig:motiondirector}
\end{figure*}

\section{Method}
MotionDuet is a diffusion-based multimodal framework that unifies text and video conditions for 3D human motion generation. 
As shown in Fig.~\ref{fig:motiondirector}, the pipeline follows a diffusion paradigm with three key steps: (1) \textbf{Video representation extraction}, in which a fine-tuned VideoMAE encoder is used to extract spatiotemporal features that capture real motion dynamics and serve as video priors. (2) \textbf{Dual-stream fusion with auto-guidance}, in which the \emph{motion–text} and \emph{video} embeddings are fused with an auto-guidance mechanism.
(3) \textbf{Multimodal distribution alignment}, in which DUET module further integrates motion-text semantics and video-grounded motion cues during diffusion training, regularized by the proposed DASH loss to align the learned motion distribution with real video statistics. Notably, with the strong regularization effect imposed by the video-conditioned training, MotionDuet retains the ability to generate high-quality and physically plausible motions using text-only prompts, significantly enhancing its practical applicability.

\subsection{Auto-Guided Dual Conditioning}

MotionDuet employs a dual-conditioning paradigm that simultaneously leverages both video and textual inputs to guide motion generation. The 3D motion sequences from the dataset are rendered through mesh skinning, followed by the generation of multi-view videos. More implementation details can be found in Appendix \ref{Video Motion Dataset}.
The video inputs provide explicit spatio-temporal trajectory control, while the textual inputs supply essential semantic guidance. 

\subsubsection{Vision and Text Conditioning}
To provide multimodal guidance, we employ two pretrained encoders: a Vision Transformer $\mathcal{E}_{\text{Vim}}$ trained based on VideoMAE \cite{wang2023videomae} for video input, and a CLIP Text Encoder $\mathcal{E}_{\text{CLIP}}$ for text prompts. Given an input video $I$ and a text prompt $\mathbf{t}$, we obtain the visual feature sequence:
\begin{equation}
    \mathbf{V} = \mathcal{E}_{\text{Vim}}(I),
\end{equation}
and the text embedding $\mathbf{T} = \mathcal{E}_{\text{CLIP}}(\mathbf{t})$, which are jointly used for conditioning downstream modules. The features from these two modalities offer complementary strengths: the text encoder  provides high-level semantic guidance and $\mathcal{E}_{\text{CLIP}}$ the video encoder $\mathcal{E}_{\text{Vim}}$ extracts rich physical motion priors. This dual-conditioning strategy ensures that the generated motions are not only aligned with the description but also physically plausible.

\subsubsection{Multimodal Fusion with Auto Guidance}
The prevailing conditional generative modeling approach CFG typically assigns static and separate guidance to each input condition during inference. 
Given the noisy motion representation $\mathbf{x}_t$ at diffusion step~$t$, the update process can be expressed as:
\begin{equation}
\begin{aligned}
\nabla \log p(\mathbf{x}_t \mid \mathbf{V}, \mathbf{T}) 
\approx \omega_{\text{v}} \nabla \log p(\mathbf{x}_t \mid \mathbf{V}) 
\\
+ \omega_{\text{t}} \nabla \log p(\mathbf{x}_t \mid \mathbf{T}),
\end{aligned}
\end{equation}
where $\omega_{\text{v}}$ and $\omega_{\text{t}}$ are manually tuned weights for the vision and text conditions, respectively.

To enable joint modeling of modalities, we employ a multimodal fusion module $\Theta_{\text{DUET}}$ to encode visual and textual inputs into a unified representation:
\begin{equation} \label{eqn:F}
\mathbf{H} = \Theta_{\text{DUET}}(\mathbf{V}, \mathbf{T}).
\end{equation}
This design treats $\mathbf{V}$ and $\mathbf{T}$ as correlated signals governed by a joint distribution \( p(\mathbf{x}_t \mid \mathbf{V}, \mathbf{T}) \), allowing the model to learn their mutual dependencies and internal balancing.

At inference time, one might apply a unified CFG weight over the fused representation $\mathbf{H}$:
\begin{equation}
\nabla \log p(\mathbf{x}_t \mid \mathbf{H}) \approx (1 \!+\! \omega) \nabla \log p(\mathbf{x}_t \mid \mathbf{H}) \!-\! \omega \nabla \log p(\mathbf{x}_t),
\end{equation}
However, such CFG-based strategies suffer from sensitivity to manually tuned weights, often leading to suboptimal balance and unstable gen. Moreover, they lack an internal correction mechanism to compensate for degraded outputs.

To address these limitations, we propose \textbf{Auto Guidance}, a novel mechanism that enables self-corrective multimodal balancing without manual weight tuning.
Inspired by the \emph{degraded model} concept introduced in~\cite{karras2024guiding}. Auto Guidance refines its own predictions by reusing the same model under varying conditioning strengths, instead of training a separate degraded network. 

Specifically, we maintain two models, $\mathcal{M}_1$ and $\mathcal{M}_2$, that share parameters but differ in conditioning intensity: $\mathcal{M}_1$ represents the clean, fully conditioned model, while $\mathcal{M}_2$ serves as its degraded counterpart with reduced conditioning. This mechanism further refines video-regularized text learning by encouraging the model to self-correct its multimodal balance during inference, where the final denoised output is computed as follows:
\begin{equation}
\begin{aligned}
M_{\text{auto}}(\mathbf{x}_t; \sigma, \mathbf{V}, \mathbf{T})
& = M_1(\mathbf{x}_t; \sigma, \mathbf{V}, \mathbf{T}) \\
 + \omega \big( 
M_1(\mathbf{x}_t; & \ \sigma, \mathbf{V}, \mathbf{T})
- M_2(\mathbf{x}_t; \sigma, \mathbf{V}, \mathbf{T})
\big),
\end{aligned}
\end{equation}
where $\omega$ is a fixed extrapolation factor. 
This formulation follows the same principle as classifier-free guidance but replaces the unconditional branch with a degraded one, enabling the model to perform self-correction using its own predictions under different conditioning levels. 
In practice, this approach stabilizes multimodal guidance and avoids manual weight tuning between modalities.

\subsection{DUET: Dual-stream Unified Encoding and Transformation}
To further enhance representational richness and mitigate potential variations in quality or informativeness across inputs, we propose DUET. It integrates four complementary branches: the Fast Fourier Transform (FFT) branch captures global periodicity and temporal regularities; the convolutional branch focuses on geometric representations and local spatial refinement; the Dynamic Mask Mechanism (DMM) adaptively selects semantically aligned and reliable features across modalities; and the residual connection helps preserve original information and stabilize the fusion process. This synergy ensures that both global structure and local details are preserved, while noisy or inconsistent inputs are effectively suppressed.


\vspace{3pt}\noindent\textbf{Fourier Branch.}
Human motion frequently exhibits periodic or quasi-periodic temporal patterns (e.g., walking, running), making frequency-domain modeling naturally suitable for capturing such dynamics. To enhance motion representation, we introduce a lightweight Fourier branch that operates in the frequency domain. Given an input feature $R$, we perform:
\begin{equation}
\mathbf{F}=\mathcal{F}^{-1}\big(W\odot \mathcal{F}(\mathbf{R})\big),
\end{equation}
where $\mathcal{F}$ is the temporal FFT, $\odot$ denotes element-wise multiplication, and $W$ is a learnable magnitude filter (we do not modify phase). This enhances periodic cues and temporal coherence.

\begin{figure*}[htbp]
    \centering
    \includegraphics[width=\linewidth]{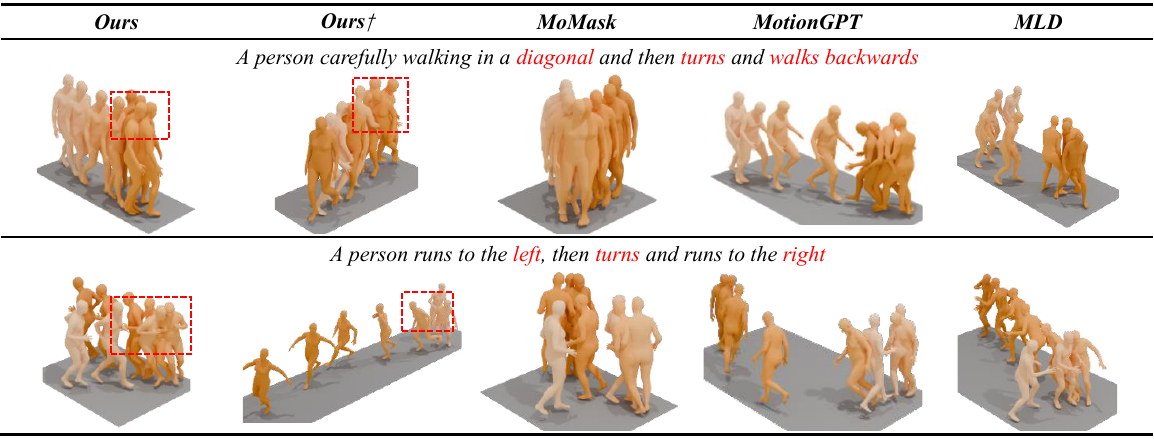}
    \caption{Qualitative results. \textbf{MotionDuet} captures motion direction and temporal coherence more accurately than prior methods, more results can be seen in Appendix~\ref{More_com_result}. MoMask uses parallel masked modeling, while MLD adopts progressive diffusion denoising. In both rows, MotionDuet achieves smoother coordination and more precise dynamics. ${\dag}$ denotes text-only inference without video guidance.}
    \label{fig:Qualitative experimental on text-to-motion.}
\end{figure*}

\vspace{3pt}\noindent\textbf{DMM.}
Video inputs may exhibit inconsistent quality across modalities, which can degrade cross-modal fusion (see Appendix \ref{Anomaly Data Analysis.}). To mitigate such variations, we introduce the DMM that adaptively preserves the modality features most aligned with the shared semantic representation.
To adaptively select the more reliable modality, we compute the distance of each modality feature to the fused representation $\mathbf{R}_{\text{fusion}}$:
\begin{equation}
d_{\text{o}} = \|\mathbf{R}_{\text{fusion}} - \mathbf{R}_{\text{o}}\|_2, \ \ 
d_{\text{b}} = \|\mathbf{R}_{\text{fusion}} - \mathbf{R}_{\text{b}}\|_2,
\end{equation}
where $\mathbf{R}_{\text{o}}$ and $\mathbf{R}_{\text{b}}$ denote the features from the motion (or “original”) and video (or “base”) branches, respectively. 
A binary mask then selects the feature that is closer to the fused representation:
\begin{equation}
\text{Mask} =
\begin{cases}
1, & \text{if } d_{\text{o}} > d_{\text{b}},\\
0, & \text{otherwise.}
\end{cases}
\end{equation}
The final fused representation is given by
\begin{equation}
\mathbf{R}_{\text{DMM}}
= \text{Mask} \cdot \mathbf{R}_{\text{o}} + (1 - \text{Mask}) \cdot \mathbf{R}_{\text{b}},
\end{equation}
and the result is concatenated with the original fusion feature $\mathbf{R}_{\text{fusion}}$ as:
\begin{equation}
\mathbf{H} = [\,\mathbf{R}_{\text{DMM}};\ \mathbf{R}_{\text{fusion}}\,].
\end{equation}

Intuitively, features more consistent with the fused representation are retained, while noisy or low-quality ones are suppressed. Note that the FFT and convolution branches operate in parallel to DMM to avoid suppressing informative regions and preserve receptive field diversity.

\vspace{3pt}\noindent\textbf{Dynamic Handling of Missing Modalities.}
MotionDuet supports both \textit{text-only} and \textit{text+video} modes without structural changes. 
When the video input is absent, its feature $\mathbf{V}$ is set to zero while keeping the text embedding $\mathbf{T}$ unchanged. 
The DUET module first constructs a joint feature $\mathbf{R}_{\text{fusion}}$ and then performs similarity-based selection through the DMM. With $\mathbf{V}$ being all-zero, its similarity becomes minimal, causing DMM to naturally route information from the motion (text-derived) branch. 
This design enables a smooth fallback to text-only conditioning without feature distortion or instability, ensuring robust generation under missing-modality scenarios.

\subsection{Auto Guidance Mechanism}
\label{Auto Guidance Mechanism}
To enable adaptive dual-conditioning in multimodal diffusion without retraining or manual tuning, we propose a lightweight guidance optimization strategy based on feature space conditional perturbation. Unlike prior works~\cite{karras2024guiding} that simulate weak conditions via input masking or model degradation, we directly perturb the fused representation $\mathbf{H}$ in feature space. This approach preserves the pretrained model weights and enables efficient guidance optimization without architecture changes, while accounting for the inherent structural differences across modalities: text embeddings are dense and semantically fragile, whereas video features exhibit high spatial-temporal redundancy and are more tolerant to perturbations.

\vspace{3pt}\noindent\textbf{Feature-Space Perturbation.}
Given the fused embedding $\mathbf{H}$ (\textit{cf.} Eq.~\eqref{eqn:F}), we simulate degraded conditions using two forms of perturbation:

\begin{itemize}
    \item \textit{Dropout Perturbation ($\mathcal{D}$):} Randomly zeros a proportion $p$ of feature dimensions:
    \begin{equation}
        \tilde{\mathbf{H}}^{(\mathcal{D})} = \text{Dropout}(\mathbf{H}; \mathcal{D}).
    \end{equation}

    \item \textit{Gaussian Noise Perturbation ($\sigma$):} Adds isotropic Gaussian noise:
    \begin{equation}
        \tilde{\mathbf{H}}^{(\sigma)} = \mathbf{H} + \epsilon, \quad \epsilon \sim \mathcal{N}(0, \sigma^2).
    \end{equation}
\end{itemize}

These operations simulate weaker or noisier conditions in latent space without altering the model architecture or requiring retraining.

\vspace{3pt}\noindent\textbf{Auto Guidance with Perturbed Features.}
Instead of relying on clean and null conditions as in classifier-free guidance, we guide the generation using a clean embedding and its degraded counterpart with controlled noise, denoted as $\tilde{\mathbf{H}}^{\text{strong}}$ and $\tilde{\mathbf{H}}^{\text{weak}}$. The final output is computed as:
\begin{equation}
\hat{\mathbf{x}}_t = (1 + \omega) \cdot \hat{\mathbf{x}}_t^{\text{strong}} - \omega \cdot \hat{\mathbf{x}}_t^{\text{weak}},
\end{equation}
where $\hat{\mathbf{x}}_t^{\text{strong}}$ and $\hat{\mathbf{x}}_t^{\text{weak}}$ are predictions conditioned on the corresponding clean and degraded features.




This formulation preserves latent-space consistency and enables gradient-free guidance without an unconditional branch, reducing sampling instability and overconfident weighting. In practice, the extrapolation factor $\omega$ is searched once via lightweight validation and fixed thereafter. 
Unlike conventional classifier-free guidance (CFG) that requires per-sample weight tuning, our approach offers stable, deployment-friendly performance across diverse conditions.

\subsection{Training Objectives}

\subsubsection{Multimodal Denoising Objective}

We adopt a denoising objective inspired by the MLD \cite{chen2023executing}, which formulates motion generation as a conditional diffusion process guided by multimodal contexts. Given a clean motion sequence $\mathbf{x}_0$ and its noisy version $\mathbf{x}_t$ at diffusion timestep $t$, the model learns to obtain the predicted latent $\hat{\mathbf{z}}_t$ using multimodal condition $\mathbf{c} = (\mathbf{V}, \mathbf{T})$ (which contains text and video embeddings extracted by frozen encoders), i.e.,
$
\hat{\mathbf{z}}_t = \mathcal{D}_\theta(\mathbf{x}_t, t, \mathbf{c}),
$
where $\mathcal{D}_\theta$ is the denoising network (a Transformer-based decoder).
The training objective minimizes the mean squared error between the predicted latent $\hat{\mathbf{z}}_t$ and the diffusion target  $\mathbf{z}_{\text{target},t}$, i.e.,
\begin{equation}
\mathcal{L}_{\text{MLD}} = \mathbb{E}_{\mathbf{x}_0, t, \boldsymbol{\epsilon}} \left[ \left\| \hat{\mathbf{z}}_t - \mathbf{z}_{\text{target},t} \right\|^2 \right].
\end{equation}

This serves as the primary supervision signal for motion generation, with additional guidance losses applied on latent representations as detailed below.

\subsubsection{Distribution-aware Training with DASH Loss}
To bridge the distributional gap between generated latent motions and real video-conditioned embeddings, we propose the \textbf{DASH} loss. 
Unlike existing objectives such as \textit{Contrastive}~\cite{radford2021learning}, \textit{Triplet}~\cite{schroff2015facenet}, or \textit{Optimal Transport} losses~\cite{peyre2019computational} that emphasize global alignment or rigid mapping, these methods often overlook fine-grained token misalignment and lack explicit structural regularization, leading to unstable training. 
DASH regularizes motion representations by enforcing both \emph{token-level similarity} and \emph{structural consistency} with video-conditioned features.

Specifically, we extract:
\begin{itemize}
    \item Motion feature tokens \(\hat{\mathbf{z}}_{t,\text{d}}\), i.e., hidden representations from the \(d\)-th layer of the denoising transformer at diffusion step \(t\), capturing intermediate structural cues. The network input includes motion latents, text, and temporal embeddings.
    \item Video reference features \(\mathbf{V}\) from the VideoMAE encoder (\textit{cf.} Eq.~(2)), encoding spatiotemporal dynamics from video inputs.
\end{itemize}

Each sample \(i \in \{1,\dots,N\}\) corresponds to a paired token \((\hat{z}_{t,\text{d},i}, v_i)\), representing aligned motion–video features within the same temporal segment.

\vspace{5pt}\noindent\textbf{Token-wise Margin Loss.}
We first align individual latent tokens to their video-conditioned counterparts using a margin-based cosine similarity loss, i.e.,
\begin{equation}
\mathcal{L}_{\text{token}} = \frac{1}{N} \sum_{i=1}^{N} \text{ReLU} \left( 1 - m_{\text{cos}} - \cos(\hat{z}_{t,\text{d},i}, v_i) \right),
\end{equation}
where $\cos(\cdot,\cdot)$ denotes the cosine similarity, and $m_{\text{cos}}$ is a predefined margin. This loss penalizes only token pairs whose similarity falls below a predefined margin, encouraging stable semantic alignment while avoiding unnecessary constraints on well-matched pairs.

\vspace{5pt}\noindent\textbf{Pairwise Structure Alignment.}
To preserve the global structure of the feature space, we introduce a structural consistency loss that aligns the pairwise similarity between token pairs within each modality, i.e.,
\begin{multline}
\mathcal{L}_{\text{pair}} = \frac{1}{N^2}\sum_{i,j=1}^{N} \text{ReLU} \big( | \cos(\hat{z}_{t,\text{d},i}, \hat{z}_{t,\text{d},j}) \\ - \cos(v_i, v_j) | - m_{\text{pair}} \big),
\end{multline}
where $m_{\text{pair}}$ is a margin threshold. This formulation encourages the relative structure of the motion latent space to mirror that of the video-conditioned embedding space.

\subsubsection{Overall Loss Formulation.}
The full DASH loss is given by a weighted sum of the two objectives, i.e.,
\begin{equation}
\mathcal{L}_{\text{DASH}} = \mathcal{L}_{\text{token}} + \mathcal{L}_{\text{pair}}.
\end{equation}

Finally, the total training loss combines the latent diffusion reconstruction objective $\mathcal{L}_{\text{MLD}}$ with our proposed alignment regularizer, i.e.,
\begin{equation}
\mathcal{L} = \mathcal{L}_{\text{MLD}} + \lambda_{\text{DASH}} \mathcal{L}_{\text{DASH}}.
\end{equation}
This distribution-aware training scheme enhances both semantic fidelity and structural coherence of the generated motions, enabling more expressive and controllable motion synthesis across modalities.


\begin{table*}[htbp]
\centering
\footnotesize
\setlength{\tabcolsep}{3pt}
\renewcommand{\arraystretch}{0.9}
\begin{tabular}{l|ccc|c|c|c|c}
\toprule
\multirow{2}{*}{\textbf{Method}} & \multicolumn{3}{c|}{\textbf{R Precision ↑}} & \multirow{2}{*}{\textbf{FID ↓}} & \multirow{2}{*}{\textbf{MM Dist ↓}} & \multirow{2}{*}{\textbf{Diversity →}} & \multirow{2}{*}{\textbf{MM ↑}} \\
 & Top 1 & Top 2 & Top 3 &  &  &  & \\
\midrule
Real             & $0.511^{\pm.003}$ & $0.703^{\pm.003}$ & $0.797^{\pm.003}$ & $0.002^{\pm.000}$ & $2.974^{\pm.008}$ & $9.503^{\pm.000}$ & --- \\ \midrule
T2M  \cite{Guo_2022_CVPR}  & $0.457^{\pm.002}$ & $0.639^{\pm.003}$ & $0.740^{\pm.004}$ & $1.067^{\pm.024}$ & $3.340^{\pm.008}$ & 
$9.188^{\pm.002}$ & $2.090^{\pm.018}$ \\
MDM \cite{tevet2022human} & $0.320^{\pm.005}$ & $0.498^{\pm.004}$ & $0.611^{\pm.004}$ & $0.544^{\pm.024}$ & $5.566^{\pm.027}$ & $9.559^{\pm.086}$ & $\best{2.799^{\pm.018}}$  \\
Fg-T2M \cite{wang2023fg} & $0.492^{\pm.002}$ & $0.683^{\pm.003}$ & $0.783^{\pm.004}$ & $0.243^{\pm.024}$ & \thirdbest{$3.109^{\pm.007}$} & $9.278^{\pm.072}$ & $1.614^{\pm.049}$  \\
MotionDiffuse  \cite{zhang2024motiondiffuse}  & $0.491^{\pm.001}$ & $0.681^{\pm.001}$ & $0.782^{\pm.001}$ & $0.630^{\pm.024}$ & $3.113^{\pm.001}$ & $9.410^{\pm.059}$ & $1.553^{\pm.042}$ \\
MotionGPT \cite{jiang2024motiongpt} & $\thirdbest{0.492^{\pm.002}}$  & $0.681^{\pm.003}$ & $0.778^{\pm.004}$ & $0.232^{\pm.024}$ &  $\secondbest{3.096^{\pm.024}}$ & 
$9.602^{\pm.071}$ & $2.008^{\pm.071}$ \\ 
CrossDiff \cite{ren2024realistic} & $0.447^{\pm.002}$ & $0.629^{\pm.003}$ & $0.730^{\pm.004}$ & $0.216^{\pm.024}$ & $3.358^{\pm.024}$ & 
 $9.577^{\pm.071}$ & $\secondbest{2.620^{\pm.071}}$  \\ 
MoMask \cite{guo2024momask} & $\best{0.504^{\pm.002}}$ & $\best{0.699^{\pm.003}}$  & $\best{0.797^{\pm.004}}$ & $\best{0.082^{\pm.024}}$  & $\best{3.050^{\pm.024}}$ & $\thirdbest{9.549^{\pm.071}}$
 & $1.241^{\pm.071}$ \\ \midrule
Baseline \cite{chen2023executing}  & $0.481^{\pm.003}$ & $0.673^{\pm.003}$ & $0.772^{\pm.002}$ & $0.473^{\pm.013}$ & $3.196^{\pm.010}$ & $9.724^{\pm.082}$ & $2.413^{\pm.079}$  \\ \midrule
Our$\dag$     & \thirdbest{$0.492^{\pm.005}$} & \thirdbest{$0.685^{\pm.003}$} & \thirdbest{$0.786^{\pm.003}$} & \thirdbest{$0.213^{\pm.024}$} & $3.176^{\pm.010}$ & \secondbest{$9.540^{\pm.071}$} & $2.464^{\pm.018}$ \\ 
Our       & $\secondbest{0.497^{\pm.003}}$ & $\secondbest{0.698^{\pm.003}}$   & $\secondbest{0.795^{\pm.003}}$ &  $\secondbest{0.179^{\pm 0.024}}$  & $3.154^{\pm.010}$  & $\best{9.532^{\pm.080}}$ &  $\thirdbest{2.496^{\pm.018}}$  \\ \midrule
Real-filtering    & $0.490^{\pm.003}$ & $0.684^{\pm.003}$ & $0.772^{\pm.002}$ & $0.002^{\pm.000}$ & $2.954^{\pm.010}$ & $9.492^{\pm.002}$ & -- \\ \midrule
Baseline-filtering    & $0.446^{\pm.003}$ & $0.628^{\pm.003}$ & $0.734^{\pm.002}$ & $0.396^{\pm.024}$ & $3.156^{\pm.010}$ & $9.710^{\pm.071}$ & $2.433^{\pm.018}$ \\ \midrule
Our-filtering$\dag$    & $0.460^{\pm.003}$ & $0.648^{\pm.003}$ & $0.754^{\pm.003}$ & $0.102^{\pm.012}$ & $3.135^{\pm.010}$ & $9.555^{\pm.071}$ & $2.860^{\pm.071}$ \\
Our-filtering   & $0.474^{\pm.003}$ & $0.668^{\pm.003}$ & $0.764^{\pm.003}$ & $0.084^{\pm.012}$ & $3.089^{\pm.010}$ & $9.527^{\pm.071}$ & $2.576^{\pm.071}$ \\
\bottomrule
\end{tabular}
\caption{
Performance comparison of various methods on the HumanML3D dataset. 
$\uparrow$ indicates higher is better, $\downarrow$ indicates lower is better, and $\rightarrow$ indicates closer is better.
'Filtering' denotes that data cleaning has been applied to the HumanML3D dataset to remove noisy or low-quality samples. \textit{$\dag$ indicates that during testing, no video was used as guidance, the motion was generated solely based on text.}
We highlight the top three results in each column with 
\textcolor{BrickRed}{\textbf{Red bold}} (best), 
\textcolor{NavyBlue}{\underline{Blue underline}} (second), and 
\textcolor{ForestGreen}{Green} (third).
}
\label{tab:comparison}
\end{table*}

\section{Experiments}
We fine-tuned the pretrained VideoMAEv2 ViT-G model on our motion video dataset (detailed in Appendix \ref{Video Motion Dataset}) using eight NVIDIA Tesla A800-80GB GPUs, with the process taking approximately one week. The VAE component was trained independently for 30 hours on a single A800-80GB GPU. Following feature extraction, all video representations were inferred and integrated into the training pipeline, which ran for about 24 hours on two NVIDIA H100-80GB GPUs. All models were trained using the AdamW optimizer with a fixed learning rate of $10^{-4}$. A batch size of 256 was used for both the VAE and diffusion training stages. The VAE was trained for 6,000 epochs, the diffusion model for 3,000 epochs, and the VideoMAE was fine-tuned for 28 epochs. Details regarding evaluation metrics and datasets are provided in Appendix \ref{Evaluation Metrics and Datasets}.

\begin{figure*}[htbp]
    \centering
    \includegraphics[width=1\linewidth]{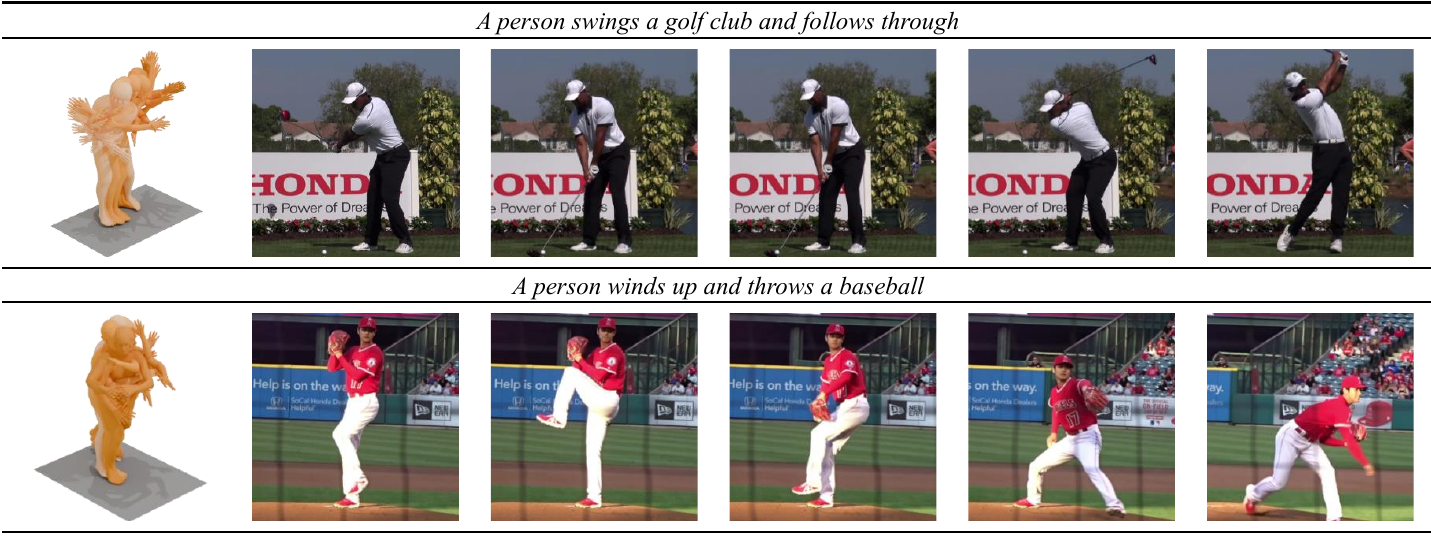}
    \caption{Qualitative results of model-generated motions for real-world videos involving complex actions. Examples include ballet spins and baseball pitching. In the golf swing sequence, the generated motion accurately captures the smooth and continuous rotation of the torso. In the baseball throwing example, the model vividly depicts the dynamic coordination between body rotation and arm extension, effectively conveying the power and fluidity of the motion. Additional qualitative results are provided in the Appendix \ref{Qualitative Evaluation of Generalization on Unseen Real-World Videos} .}
    \label{fig:real-world videos involving complex actions_1}
\end{figure*}
\subsection{Evaluation on Motion Generation}
We evaluate MotionDuet on the HumanML3D~\cite{guo2022generating} dataset following~\cite{chen2023executing}. 
As shown in Table~\ref{tab:comparison}, our model performs strongly across all metrics, achieving an R@3 of $0.795$ and a low FID of $0.179$, indicating high realism. 
Diversity and MM scores also improve consistently, validating the model’s effectiveness in generating accurate and varied text-conditioned motions. 
Qualitative examples are provided in Fig.~\ref{fig:Qualitative experimental on text-to-motion.} and Appendix~\ref{Qualitative Experiment}, with additional results on unseen real-world videos in Fig.~\ref{fig:real-world videos involving complex actions_1}.

Although our FID and R@3 scores are slightly lower than those of MoMask, this is primarily due to the introduction of video-based features during training. While these features are not derived from real-world videos, they belong to a distinct video modality whose distribution differs from that of motion representations. This inherent modality gap can affect metrics such as FID and R@3, which are sensitive to distributional alignment, but it does not accurately reflect perceptual motion quality. As shown in Fig.~\ref{fig:real-world videos involving complex actions_1}, motions generated by \textbf{MotionDuet} exhibit comparable visual fidelity and notably stronger directional and semantic control. Overall, our framework achieves a balanced trade-off between quantitative metrics and qualitative fidelity, providing enhanced controllability and alignment in text-conditioned motion generation.

\begin{table}[htbp]
\centering
\footnotesize
\setlength{\tabcolsep}{4pt}  
\renewcommand{\arraystretch}{0.6}
\begin{tabular}{l|c|c|c}
\toprule
\textbf{Method} & \textbf{R@3 ↑} & \textbf{FID ↓} & \textbf{MM Dist ↓} \\
\midrule
Real-filtering    & $0.772^{\pm.002}$ & $0.002^{\pm.000}$ & $2.954^{\pm.010}$ \\ \midrule
Element-Wise Add       & $0.747^{\pm.003}$ & $0.168^{\pm.012}$ & $3.388^{\pm.010}$ \\
\hspace*{1em}+ DMM         & $0.750^{\pm.003}$ & $0.204^{\pm.012}$ & $3.256^{\pm.010}$ \\ 
\hspace*{1em}+ FFT & $0.750^{\pm.003}$ & $0.163^{\pm.012}$ & $3.178^{\pm.010}$ \\ 
\hspace*{1em}+ Identity     & $0.752^{\pm.003}$ & $0.147^{\pm.012}$ & $3.124^{\pm.010}$ \\ 
\hspace*{1em}+ Conv & \bm{$0.755^{\pm.003}$} & \bm{$0.101^{\pm.024}$} & \bm{$3.087^{\pm.010}$} \\
\bottomrule
\end{tabular}
\caption{Performance comparison of multimodal fusion strategies. The top results in each column are highlighted with 
\textbf{bold}. More feature fusion comparison results are shown in Appendix \ref{Supplementary Data on Multimodal Fusion Strategies}.}
\label{tab:fusion_strategies}
\end{table}
\subsection{Ablation Study}


\vspace{3pt}\noindent\textbf{Evaluation on Multimodal Fusion Strategies.}
We compare multiple multimodal fusion strategies on the filtered HumanML3D dataset, removing the DASH Loss to isolate fusion effects (Table~\ref{tab:fusion_strategies}). Among standard baselines (e.g., concatenation, cross-attention, and element-wise operations), element-wise addition consistently delivers the most stable and competitive performance. Building on this observation, we enhance element-wise fusion with four parallel complementary branches, forming our DUET module. DUET markedly improves integration quality. Full details of fusion variants and search strategies are included in Appendix \ref{Supplementary Data on Multimodal Fusion Strategies}.



\vspace{3pt}\noindent\textbf{Evaluation on Each Component.}
We conduct an ablation study to evaluate each component (Table~\ref{tab:ablation_no_mm}). 
After constructing and cleaning the video-based motion dataset (Appendix~\ref{Automated Vidoe Data Cleaning}), re-evaluating the baseline already yields notable metric gains. 

\begin{table}[htbp]
\centering
\footnotesize
\setlength{\tabcolsep}{4pt}  
\renewcommand{\arraystretch}{0.7}
\begin{tabular}{l|c|c|c}
\toprule
\textbf{} & \textbf{R@3 ↑} & \textbf{FID ↓} & \textbf{MM Dist ↓} \\
\midrule
Real             & $0.797^{\pm.003}$ & $0.002^{\pm.000}$ & $2.974^{\pm.008}$ \\
\midrule
Baseline         & $0.772^{\pm.002}$ & $0.473^{\pm.024}$ & $3.196^{\pm.010}$ \\
+ Filtering      & $0.734^{\pm.002}$ & $0.396^{\pm.024}$ & $3.156^{\pm.010}$ \\
\midrule
Real-filtering   & $0.772^{\pm.002}$ & $0.002^{\pm.000}$ & $2.954^{\pm.010}$ \\
\midrule
+ Video          & $0.742^{\pm.003}$ & $0.192^{\pm.012}$ & $3.296^{\pm.010}$ \\
+ DUET           & $0.755^{\pm.003}$ & $0.101^{\pm.024}$ & \bm{$3.087^{\pm.010}$} \\
+ DASH Loss          & \bm{$0.764^{\pm.003}$} & \bm{$0.084^{\pm.012}$} & $3.089^{\pm.010}$ \\
\bottomrule
\end{tabular}
\caption{The effectiveness of each module has been validated, with the best results per column highlighted in \textbf{bold}. More loss comparison results are shown in Appendix \ref{Evaluation on each component}.}
\label{tab:ablation_no_mm}
\end{table}

\subsubsection{Evaluation on Encoder Tuning and Model Scale}

To study the impact of encoder training and capacity on motion generation, we ablate different VideoMAEv2 backbones (Table~\ref{tab:video_encoder_1}). 
We compare a zero-shot and a fine-tuned ViT-G encoder, along with a distilled ViT-B encoder, to highlight the impact of fine-tuning and model scale on both performance and efficiency.

\begin{table}[htbp]
\centering
\footnotesize
\renewcommand{\arraystretch}{1}
\begin{tabular}{l|c|c|c}
\toprule
\textbf{Method} & \textbf{R@3 ↑} & \textbf{FID ↓} & \textbf{MM Dist ↓} \\
\midrule
Real                       & $0.797^{\pm.003}$ & $0.002^{\pm.000}$ & $2.974^{\pm.008}$ \\
\midrule
MLD (Baseline)             & $0.772^{\pm.002}$ & $0.473^{\pm.013}$ & $3.196^{\pm.010}$ \\
ViT-G (fine-tuned)         & $0.795^{\pm.003}$ & $0.179^{\pm.024}$ & $3.154^{\pm.010}$ \\
ViT-G (frozen)             & $0.751^{\pm.003}$ & $0.238^{\pm.024}$ & $3.334^{\pm.010}$ \\
ViT-B (fine-tuned)         & $0.782^{\pm.003}$ & $0.182^{\pm.012}$ & $3.178^{\pm.010}$ \\
\bottomrule
\end{tabular}
\caption{Comparison of video encoders on HumanML3D. More results are shown in Appendix \ref{Quantitative Evaluation of the Video Encoders}.}
\label{tab:video_encoder_1}
\end{table}

\subsubsection{Evaluation on Auto-Guidance Mechanism}



Automatic guidance enhances generation by comparing predictions from a strong and a deliberately weakened model, amplifying updates when their outputs diverge~\cite{karras2024guiding}. 
Under multimodal settings, we evaluate two key factors: the modality weight~$\omega$ and the perturbation strategy used to construct the weaker model.

We investigate two degradation types:
\begin{itemize}
    \item \textit{Dropout-based} ($\mathcal{D}_1$, $\mathcal{D}_2$): applying $5\%$ and $10\%$ feature dropout to emulate a weaker model.
    \item \textit{Noise-based} ($\epsilon_1$, $\epsilon_2$): adding Gaussian noise with increasing strength to corrupt input embeddings.
\end{itemize}

For each case, $\omega$ is swept to identify the optimal guidance strength (Table~\ref{tab:Hyperparameters_2}). 
Dropout-based degradation provides more stable and consistent gains than noise injection, confirming its effectiveness for multimodal auto guidance. 
Additional analysis is provided in Appendix~\ref{Study on Auto-Guidance Mechanism Weights}.

\begin{table}[htbp]
\centering
\footnotesize
\renewcommand{\arraystretch}{1}
\begin{tabular}{cc|c|c|c}
\toprule
\textbf{Setting} & \textbf{$\omega$} & \textbf{R@3 ↑} & \textbf{FID ↓} & \textbf{MM Dist ↓} \\
\midrule
Real-filtering & -- & $0.772^{\pm.002}$ & $0.002^{\pm.000}$ & $2.954^{\pm.010}$ \\
\midrule

\multirow{1}{*}{\textbf{$\mathcal{D}_1$ 5\%}} 
& 1.25 & $0.764^{\pm.003}$ & $0.084^{\pm.012}$ & $3.089^{\pm.010}$ \\
\midrule

\multirow{1}{*}{\textbf{$\mathcal{D}_2$ 10\%}} 
& 0.75 & $0.755^{\pm.004}$ & $0.102^{\pm.020}$ & $3.090^{\pm.012}$ \\
\midrule

\multirow{1}{*}{\textbf{$\epsilon_1$ 5\%}} 
& 1.25 & $0.743^{\pm.003}$ & $0.101^{\pm.020}$ & $3.088^{\pm.011}$ \\
\midrule

\multirow{1}{*}{\textbf{$\epsilon_1$ 10\%}} 
& 1.00 & $0.737^{\pm.004}$ & $0.134^{\pm.022}$ & $3.103^{\pm.012}$ \\

\midrule

\multirow{1}{*}{\textbf{CFG}} 
& 6.5 & $0.737^{\pm.004}$ & $0.133^{\pm.023}$ & $3.088^{\pm.012}$ \\

\bottomrule
\end{tabular}
\caption{Parameter study for $\omega$ and dropout. Only core metrics reported. More grid searching results are shown in Appendix \ref{Study on Auto-Guidance Mechanism Weights}.}
\label{tab:Hyperparameters_2}
\end{table}

\section{Conclusion}
In summary, we present \textbf{MotionDuet}, a dual-conditioned motion generation framework that regularizes text-based motion learning with video supervision. By combining video-grounded spatiotemporal precision with text-driven semantic alignment, MotionDuet effectively bridges the distribution gap between synthesized and real human dynamics. Our design integrates the DUET fusion module, the DASH distribution-aware loss, and an auto-guidance mechanism to jointly enhance structural coherence, controllability, and realism. Extensive experiments demonstrate that MotionDuet consistently surpasses strong baselines, validating the effectiveness of video-regularized text learning for multimodal human motion generation.


{
    \small
    \bibliographystyle{IEEEtran}
    \bibliography{main}
}

\onecolumn
\appendix
\begin{center}
\large
  \textbf{Appendix}
\end{center}

\section{Evaluation Metrics and Datasets}
\label{Evaluation Metrics and Datasets}
\subsection{Evaluation Metrics}
\textbf{(1) Motion Quality}: Fréchet Inception Distance (FID) quantifies the similarity between generated and real motions in feature space; lower scores indicate better quality. \textbf{(2) Generation Diversity}: Diversity (DIV) measures variation across generated motions \cite{guo2022generating}, while Multimodality (MM) evaluates diversity for multiple generations from identical inputs. \textbf{(3) Conditional Matching}: Motion Retrieval Accuracy (R Accuracy) computes Top 1/2/3 matches between text and motion, and Multimodal Distance (MM Dist) measures text-motion feature similarity \cite{guo2022generating}. 

\subsection{Datasets}
HumanML3D \cite{Guo_2022_CVPR}, combining HumanAct12 \cite{guo2020action2motion} and AMASS \cite{AMASS:2019}, features 14,616 motions spanning daily tasks, sports, acrobatics, and artistic performances. Annotated via Amazon MTurk, each clip includes $3$-$4$ sentences, downsampled to $20$ fps, lasting $2$-$10$ s (avg. $7.1$ s), totaling $28.59$ hours. The dataset has $44,970$ descriptions averaging $12$ words each from a vocabulary of $5,371$ unique words.

\section{Qualitative Experiment}
\label{Qualitative Experiment}
\subsection{Qualitative Evaluation on Text to Motion Generation}
\begin{figure}[htbp]
    \centering
    \includegraphics[width=1\linewidth]{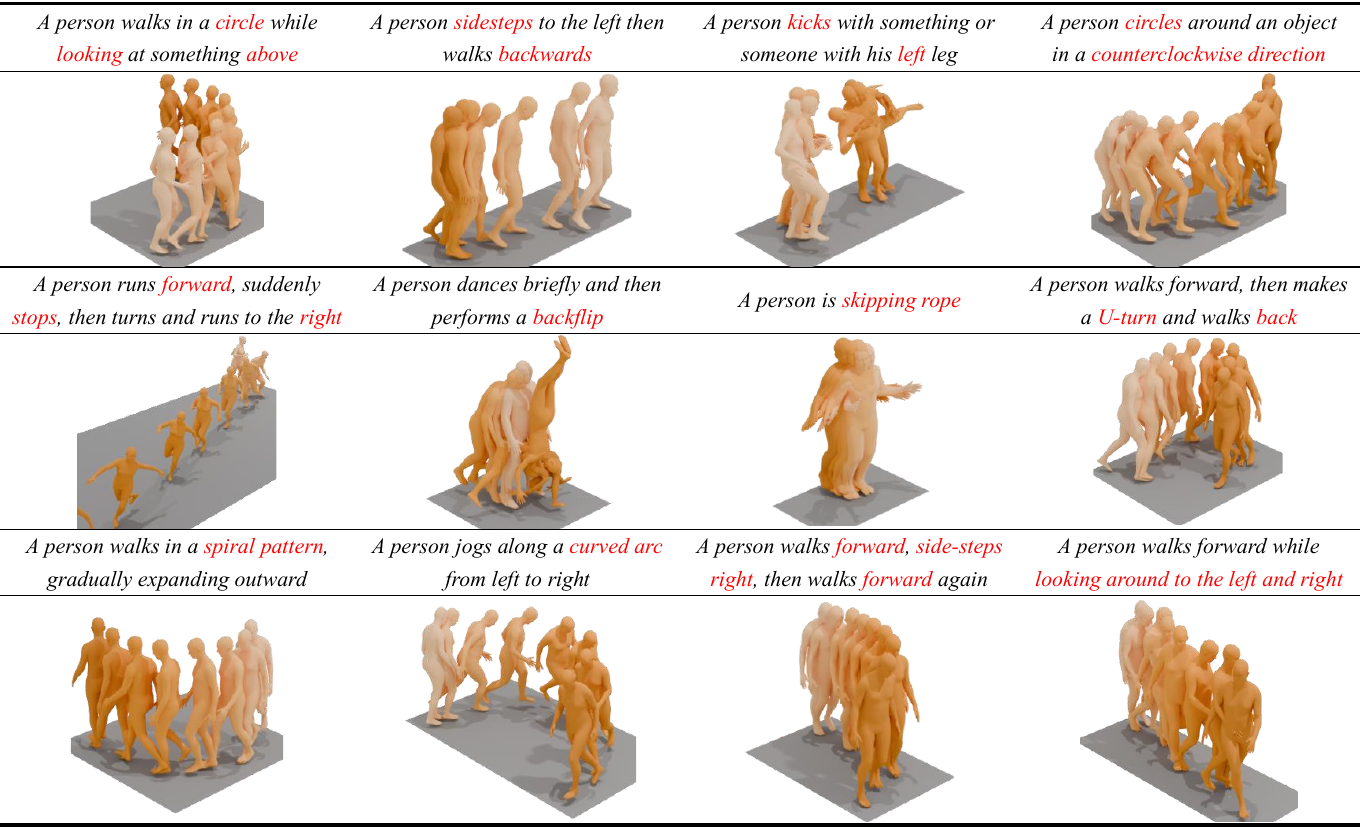}
    \caption{Qualitative experimental results.  These examples cover a variety of challenging textual descriptions, involving complex action compositions and directional changes. MotionDuet is capable of generating motion sequences at a rate of approximately 199.61 poses per second during inference.}
    \label{fig:Qualitative Experimental Results}
\end{figure}

We present a series of visualized motion results generated by our method to further evaluate its performance in real-world generation scenarios. These examples cover a variety of challenging textual descriptions, involving complex action compositions and directional changes, see Fig. 
 \ref{fig:Qualitative Experimental Results}. By directly comparing the input text with the corresponding generated motion sequences, we can clearly observe the model’s capability to understand semantic intent, capture motion details, and maintain temporal coherence. These visual results not only demonstrate the model’s precise response to natural language instructions but also highlight its strength in producing natural, coherent, and semantically consistent human motions.


\subsection{Qualitative Ablation on Video-Guided Motion Generation}
\label{More_com_result}
\begin{figure}[htbp]
    \centering
    \includegraphics[width=\linewidth]{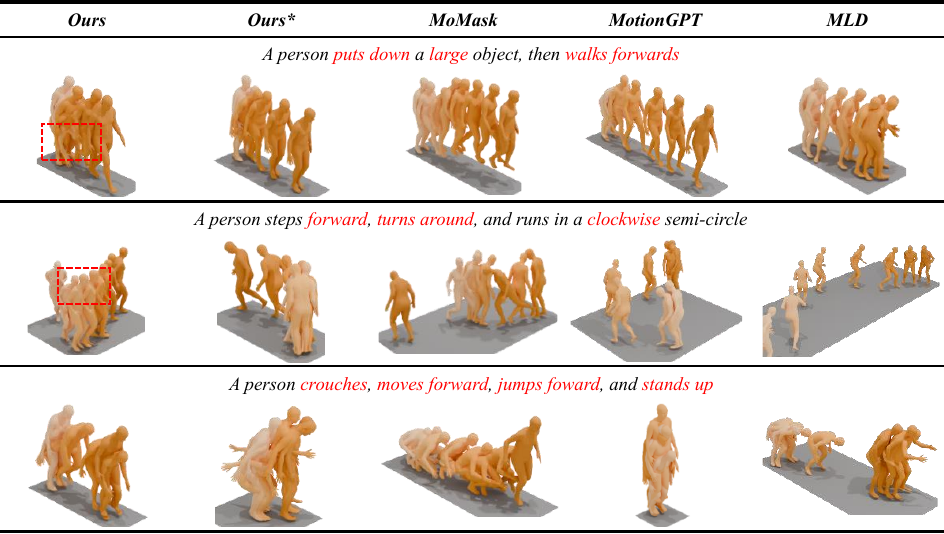}
    \caption{Comparison of qualitative experimental results. We conduct a qualitative comparison with three methods: MoMask, MotionGPT, and MLD. Compared to previous methods, our model generates more realistic and coherent motions, with better alignment to fine-grained language instructions such as “puts down a large object”, “turn around”, and “crouches and jumps forward”. Our*  denotes an ablation variant in which video inputs are excluded during training to validate their contribution to model performance. As video information is unavailable in this setting, the DASH Loss is removed accordingly, while the other components of the DUET module, excluding DMM, are retained. }
    \label{fig:Comparison of Qualitative Experimental Results}
\end{figure}

To deepen this comparison and isolate the contribution of video inputs, we also perform an ablation study in which video inputs are excluded during training. As a result, the DASH Loss is removed due to its reliance on video information, while the remaining components of the DUET module, except for DMM, are preserved to ensure a consistent and fair evaluation. In addition, we conduct qualitative evaluations of the generated motion sequences across a diverse set of textual prompts to further assess the effectiveness of our proposed method. As shown in Fig. \ref{fig:Comparison of Qualitative Experimental Results}, our model excels at generating realistic and semantically aligned human motions in response to complex natural language descriptions.

Compared to baseline models, our approach demonstrates superior physical plausibility and motion continuity, particularly in managing transitions between distinct motion primitives (e.g., turning, running, or crouching). These results underscore the model’s ability to produce context-aware, text-consistent motions in scenarios demanding precise temporal ordering and stylistic fidelity. Overall, these qualitative examples highlight our method’s exceptional ability to capture both high-level semantic intent and fine-grained motion dynamics.

\subsection{Qualitative Evaluation of Generalization on Unseen Real-World Videos}
\label{Qualitative Evaluation of Generalization on Unseen Real-World Videos}
\begin{figure}[htbp]
    \centering
    \includegraphics[width=1\linewidth]{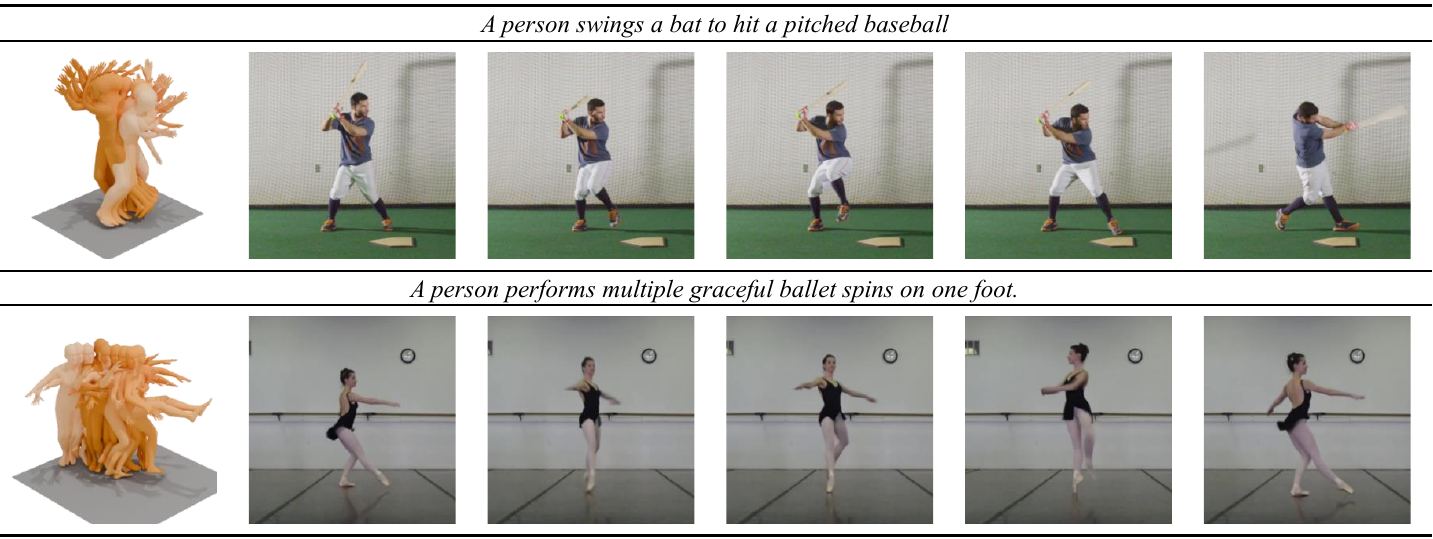}
    \caption{Qualitative results of model-generated motions for real-world videos involving complex actions. The examples include ballet spins, baseball pitching, hitting an incoming baseball with a bat, and golf swings. Although the model was never exposed to these specific videos during training, it successfully produces semantically consistent and physically plausible motions, demonstrating its ability to generalize to unseen real-world inputs.}
    \label{fig:real-world videos involving complex actions}
\end{figure}

To rigorously evaluate the model’s real-world applicability and generalization ability, we select real-life videos from the reference \cite{dong2020motion}, none of which appear during training or are included in the dataset. These videos are preprocessed and carefully trimmed into the input format required by our model. The selected samples feature several representative and high-difficulty actions, such as ballet spins, baseball pitching, hitting an incoming baseball with a bat, and golf swings (see Fig. \ref{fig:real-world videos involving complex actions_1} and Fig.~\ref{fig:real-world videos involving complex actions}). This evaluation serves as a strong qualitative test of the model’s ability to handle complex real-world motion scenarios.

When simulating the action of hitting a baseball with a bat, the model successfully reproduces the complete process, including lifting the bat overhead, swinging it clockwise, and making contact with the ball. In the case of the ballet turn, the model demonstrates a clear understanding of the structural subtleties of the movement, accurately portraying the dancer’s posture as they balance on one foot and rotate their body with grace. These results collectively highlight the model’s capability to generate realistic, coherent, and diverse human motions across a wide range of complex actions.

In the table, the first column presents the motion sequences generated by our model. The accompanying text above each sequence is a manually written description based on the corresponding video content. The remaining five columns display the reference frames, which are sampled from the original real-life video at evenly spaced intervals.

\section{Automated Video Data Cleaning}
\label{Automated Vidoe Data Cleaning}
To ensure high-quality data input for downstream motion analysis tasks, we implement a robust data cleaning algorithm that filters out erroneous or low-quality video samples based on human body orientation consistency. The method utilizes pose landmarks extracted via MediaPipe and evaluates the subject's orientation through a series of geometric and kinematic criteria. The key components of the cleaning algorithm are outlined as follows:

Let a video sample $V = \{I_t\}_{t=1}^T$ consist of $T$ frames. For computational efficiency, we sample a fixed subset of frames $\mathcal{F} = \{ I_{t_i} \mid i = 1, 2, \dots, N \}$ where $N \ll T$ using a uniform sampling strategy. Each frame $I_{t_i}$ is processed by a pose estimator to extract a set of 3D landmarks $\mathbf{L}_{t_i} \in \mathbb{R}^{J \times 3}$, where $J$ is the number of body joints.

\subsection{Back-Face Consistency}
Let $\vec{v}_{\text{back}} = \mathbf{L}_{\text{RShoulder}} - \mathbf{L}_{\text{LShoulder}}$
and $\vec{v}_{\text{hip}} = \mathbf{L}_{\text{RHip}} - \mathbf{L}_{\text{LHip}}$.
The body orientation vector is defined as
$$
\vec{v}_{\text{body}} = \frac{1}{2} (\vec{v}_{\text{back}} + \vec{v}_{\text{hip}}).
$$
We also define the face direction vector as
$$
\vec{v}_{\text{face}} = \mathbf{L}_{\text{Nose}} - \mathbf{L}_{\text{MidShoulder}},
$$
where $\mathbf{L}_{\text{MidShoulder}} = \frac{1}{2} (\mathbf{L}_{\text{LShoulder}} + \mathbf{L}_{\text{RShoulder}})$.
The body-face angle $\theta_{\text{bf}}$ is computed as
$$
\theta_{\text{bf}} = \arccos \left( \frac{\vec{v}_{\text{body}} \cdot \vec{v}_{\text{face}}}{\|\vec{v}_{\text{body}}\| \cdot \|\vec{v}_{\text{face}}\|} \right).
$$
A frame is valid if $\theta_{\text{bf}} \leq \theta_0$, where $\theta_0 = 20^\circ$.

\subsection{Head Pose Constraint}
Let $\vec{v}_{\text{head}} = \mathbf{L}_{\text{Nose}} - \mathbf{L}_{\text{MidShoulder}}$.
We constrain the head tilt angle $\theta_{\text{head}}$ against the vertical axis:
$$
\theta_{\text{head}} = \arccos \left( \frac{\vec{v}_{\text{head}} \cdot \vec{e}_y}{\|\vec{v}_{\text{head}}\|} \right).
$$
A frame is valid if $\theta_{\text{head}} \leq \theta_1$, with $\theta_1 = 30^\circ$, and $\vec{e}_y$ is the global vertical axis.

\subsection{Foot-Knee Direction Alignment}

To ensure the plausibility of gait or standing postures, we constrain the angle between the hip-to-knee vector and the ankle-to-foot vector. For each leg side $s \in \{\text{Left}, \text{Right}\}$, we define the foot-knee angle as:

$$
\theta_{\text{fk}}^{(s)} = \angle \left( \mathbf{L}_{\text{Hip}}^{(s)} - \mathbf{L}_{\text{Knee}}^{(s)},\ \mathbf{L}_{\text{Foot}}^{(s)} - \mathbf{L}_{\text{Ankle}}^{(s)} \right),
$$
where $\mathbf{L}_{\text{Hip}}^{(s)}$, $\mathbf{L}_{\text{Knee}}^{(s)}$, $\mathbf{L}_{\text{Ankle}}^{(s)}$, and $\mathbf{L}_{\text{Foot}}^{(s)}$ are the coordinates of the respective joints on side $s$.

The frame is considered \emph{valid} with respect to foot-knee alignment if:

$$
\theta_{\text{fk}}^{(s)} \in [75^\circ,\ 180^\circ], \quad \forall s \in \{\text{Left}, \text{Right}\}
$$

This constraint effectively filters out frames exhibiting unnatural foot twisting or anatomical inconsistencies, which often arise from pose tracking failures or annotation noise.


\subsection{Frame Validity and Video Filtering}

A frame $I_{t_i}$ is marked as \textit{valid} if it satisfies all of the following four constraints:
(1) back-face consistency,
(2) head pose constraint, and
(3) foot-knee direction alignment.

Let $B_i$, $H_i$, and $F_i$ denote Boolean indicators (1 if satisfied, 0 otherwise) for these three conditions on frame $i$.
We define the overall video validity score as:

$$
P(v) = \frac{1}{N} \sum_{i=1}^{N} \mathbb{I}(B_i \land H_i \land F_i)
$$

A video is considered valid if:

$$
P(v) \geq \rho,
$$
where $\rho = 0.7$ is the minimum acceptable ratio of valid frames.

To construct the cleaned validation dataset, we apply this automated filtering process to all raw videos.
Each video is uniformly sampled into $N = 12$ frame, pose landmarks are extracted via MediaPipe, and only videos passing the threshold are retained.
This ensures that downstream models are trained on reliable, consistent human motion data, free from noisy or erroneous poses.




\begin{table*}[h]
\centering
\small
\renewcommand{\arraystretch}{1}
\begin{tabular}{c|ccc|c|c|c|c}
\toprule
\multirow{2}{*}{\textbf{$\lambda_{\text{DASH}}$}} & \multicolumn{3}{c|}{\textbf{R Precision ↑}} & \multirow{2}{*}{\textbf{FID ↓}} & \multirow{2}{*}{\textbf{MM Dist ↓}} & \multirow{2}{*}{\textbf{Diversity →}} & \multirow{2}{*}{\textbf{MM ↑}} \\
 & Top 1 & Top 2 & Top 3 &  &  &  & \\
\midrule
Real-filtering    & $0.490^{\pm.003}$ & $0.684^{\pm.003}$ & $0.772^{\pm.002}$ & $0.002^{\pm.000}$ & $2.954^{\pm.010}$ & $9.492^{\pm.002}$ & -- \\ \midrule
0.1        & $0.474^{\pm.003}$ & $0.668^{\pm.003}$ & $0.764^{\pm.003}$ & $0.084^{\pm.012}$ & $3.089^{\pm.010}$ & $9.527^{\pm.071}$ & $2.576^{\pm.071}$ \\
0.3       & $0.466^{\pm.003}$ & $0.657^{\pm.003}$ & $0.752^{\pm.002}$ & $0.143^{\pm.024}$ & $3.169^{\pm.010}$ & $9.532^{\pm.071}$ & $2.453^{\pm.018}$ \\
0.5    & $0.469^{\pm.003}$ & $0.645^{\pm.003}$ & $0.745^{\pm.002}$ & $0.186^{\pm.024}$ & $3.280^{\pm.010}$ & $9.810^{\pm.071}$ &$2.456^{\pm.018}$ \\ 
0.7    & $0.452^{\pm.003}$ & $0.647^{\pm.003}$ & $0.743^{\pm.002}$ & $0.237^{\pm.024}$ & $3.311^{\pm.010}$ & $9.314^{\pm.071}$ & $2.412^{\pm.018}$ \\ 
0.9            & $0.443^{\pm.003}$ & $0.632^{\pm.003}$ & $0.734^{\pm.003}$ & $0.294^{\pm.012}$ & $3.324^{\pm.010}$ & $9.277^{\pm.071}$ & $2.427^{\pm.018}$ \\
1    & $0.433^{\pm.003}$ & $0.638^{\pm.003}$ & $0.732^{\pm.003}$ & $0.427^{\pm.024}$ & $3.322^{\pm.010}$ & $9.212^{\pm.071}$ & $2.563^{\pm.018}$ \\
50    & $0.345^{\pm.003}$ & $0.525^{\pm.003}$ & $0.635^{\pm.003}$ & $1.438^{\pm.024}$ & $3.997^{\pm.010}$ & $8.653^{\pm.071}$ & $2.672^{\pm.018}$ \\
100   & $0.310^{\pm.003}$ & $0.474^{\pm.003}$ & $0.600^{\pm.003}$ & $2.500^{\pm.012}$ & $4.275^{\pm.010}$ & $8.731^{\pm.071}$ & $2.654^{\pm.018}$ \\
200   & $0.159^{\pm.003}$ & $0.278^{\pm.003}$ & $0.369^{\pm.003}$ & $8.676^{\pm.012}$ & $5.660^{\pm.010}$ & $7.369^{\pm.071}$ & $2.684^{\pm.018}$ \\
300   & $0.039^{\pm.003}$ & $0.058^{\pm.003}$ & $0.099^{\pm.003}$ & $14.676^{\pm.012}$ & $7.320^{\pm.010}$ & $5.832^{\pm.071}$ & $2.953^{\pm.018}$ \\
\bottomrule
\end{tabular}
\caption{Parameter Study on $\lambda_{\text{DASH}}$. $\uparrow$ indicates higher is better, and $\downarrow$ indicates lower is better.}
\label{tab:Hyperparameters_1}
\end{table*}

\section{Additional Experiments}
\subsection{Evaluation of Hyperparameters $\lambda_{\text{DASH}}$}
In this section, we first conduct a detailed analysis and discussion on the range of values for the hyperparameter $\lambda_{\text{DASH}}$, aiming to understand its influence on model performance, see Table~\ref{tab:Hyperparameters_1}. Experimental results reveal a clear trend: while introducing the DASH loss with a moderate weight can effectively improve the quality and consistency of motion generation, setting $\lambda_{\text{DASH}}$ too high leads to a noticeable performance degradation. This is likely because an excessively strong DASH loss may overpower other learning signals, causing the model to overfit to the video features and thereby reducing its generalization ability, especially when video inputs are unavailable at inference time.

\subsection{Study on Auto-Guidance Mechanism Weights $\omega$}
\label{Study on Auto-Guidance Mechanism Weights}
Automatic guidance identifies and corrects potential errors by measuring the discrepancy between the predictions of a strong model and a weaker one, thereby amplifying adjustments in a more favorable direction. When the two models produce similar outputs, the perturbation is negligible; however, when they diverge, the difference serves as an approximate signal toward a better sample distribution \cite{karras2024guiding}. To investigate the effectiveness of our Auto Guidance under multimodal settings, we conduct an ablation study on two key factors: the modality-specific influence weights $\omega$ and the perturbation strategies—dropout and input noise. Specifically, we evaluate three groups of settings:

\begin{itemize}
    \item \textbf{Dropout-only} configurations: $\mathcal{D}_1$ and $\mathcal{D}_2$ represent feature-level dropout rates (e.g., $5\%$ and $10\%$) applied post-hoc to the base model. The guidance model operates using these degraded features to mimic a weaker model variant.
    
    \item \textbf{Noise-only} configurations: $\epsilon_1$ and $\epsilon_2$ indicate different levels of Gaussian noise (e.g., standard deviation increments of $5\%$ and $10\%$) added to the input embeddings. This simulates corrupted conditions to encourage robust generation.
    
\end{itemize}

\begin{table*}[htbp]
\centering
\small
\renewcommand{\arraystretch}{1}
\begin{tabular}{cc|ccc|c|c|c|c}
\toprule
\multirow{2}{*}{\textbf{$\mathcal{D}_1$}}& \multirow{2}{*}{\textbf{$\omega$}} & \multicolumn{3}{c|}{\textbf{R Precision ↑}} & \multirow{2}{*}{\textbf{FID ↓}} & \multirow{2}{*}{\textbf{MM Dist ↓}} & \multirow{2}{*}{\textbf{Diversity →}} & \multirow{2}{*}{\textbf{MM ↑}} \\
& & Top 1 & Top 2 & Top 3 &  &  &  & \\
\midrule
Real-filtering   & -- & $0.490^{\pm.003}$ & $0.684^{\pm.003}$ & $0.772^{\pm.002}$ & $0.002^{\pm.000}$ & $2.954^{\pm.010}$ & $9.492^{\pm.002}$ & -- \\ \midrule
\multirow{5}{*}{\textbf{5\%}}
& 0.75  & $0.462^{\pm.005}$ & $0.651^{\pm.006}$ & $0.742^{\pm.005}$ & $0.121^{\pm.022}$ & $3.095^{\pm.014}$ & $9.320^{\pm.077}$ & $2.543^{\pm.065}$ \\
& 1.00 & $0.469^{\pm.004}$ & $0.657^{\pm.004}$ & $0.744^{\pm.003}$ & $0.142^{\pm.018}$ & $3.082^{\pm.009}$ & $9.355^{\pm.069}$ & $2.580^{\pm.073}$ \\
& 1.25  & $0.474^{\pm.003}$ & $0.668^{\pm.003}$ & $0.764^{\pm.003}$ & $0.084^{\pm.012}$ & $3.089^{\pm.010}$ & $9.527^{\pm.071}$ & $2.576^{\pm.071}$ \\
& 1.50  & $0.463^{\pm.005}$ & $0.654^{\pm.005}$ & $0.745^{\pm.005}$ & $0.102^{\pm.024}$ & $3.100^{\pm.015}$ & $9.310^{\pm.073}$ & $2.598^{\pm.068}$ \\
& 1.75 & $0.469^{\pm.004}$ & $0.657^{\pm.004}$ & $0.749^{\pm.003}$ & $0.097^{\pm.018}$ & $3.082^{\pm.009}$ & $9.355^{\pm.069}$ & $2.513^{\pm.073}$ \\
\midrule

\multirow{2}{*}{\textbf{$\mathcal{D}_2$}}& \multirow{2}{*}{\textbf{$\omega$}} & \multicolumn{3}{c|}{\textbf{R Precision ↑}} & \multirow{2}{*}{\textbf{FID ↓}} & \multirow{2}{*}{\textbf{MM Dist ↓}} & \multirow{2}{*}{\textbf{Diversity →}} & \multirow{2}{*}{\textbf{MM ↑}} \\
& & Top 1 & Top 2 & Top 3 &  &  &  & \\
\midrule

\multirow{5}{*}{\textbf{10\%}} 
& 0.75  & $0.468^{\pm.004}$ & $0.662^{\pm.003}$ & $0.755^{\pm.004}$ & $0.102^{\pm.020}$ & $3.090^{\pm.012}$ & $9.480^{\pm.075}$ & $2.572^{\pm.068}$ \\
& 1.0 & $0.473^{\pm.003}$ & $0.666^{\pm.003}$ & $0.758^{\pm.003}$ & $0.132^{\pm.019}$ & $3.082^{\pm.011}$ & $9.503^{\pm.070}$ & $2.579^{\pm.071}$ \\
& 1.25  & $0.474^{\pm.003}$ & $0.668^{\pm.003}$ & $0.760^{\pm.003}$ & $0.153^{\pm.018}$ & $3.089^{\pm.010}$ & $9.510^{\pm.072}$ & $2.580^{\pm.069}$ \\
& 1.50 & $0.471^{\pm.004}$ & $0.663^{\pm.003}$ & $0.756^{\pm.004}$ & $0.113^{\pm.022}$ & $3.088^{\pm.011}$ & $9.495^{\pm.071}$ & $2.575^{\pm.070}$ \\
& 1.75  & $0.469^{\pm.003}$ & $0.661^{\pm.004}$ & $0.753^{\pm.003}$ & $0.323^{\pm.021}$ & $3.085^{\pm.010}$ & $9.485^{\pm.073}$ & $2.570^{\pm.068}$ \\
\midrule

\multirow{2}{*}{\textbf{$\epsilon_1$}}& \multirow{2}{*}{\textbf{$\omega$}} & \multicolumn{3}{c|}{\textbf{R Precision ↑}} & \multirow{2}{*}{\textbf{FID ↓}} & \multirow{2}{*}{\textbf{MM Dist ↓}} & \multirow{2}{*}{\textbf{Diversity →}} & \multirow{2}{*}{\textbf{MM ↑}} \\
& & Top 1 & Top 2 & Top 3 &  &  &  & \\
\midrule

\multirow{5}{*}{\textbf{5\%}} 
& 0.75  & $0.458^{\pm.005}$ & $0.645^{\pm.005}$ & $0.735^{\pm.004}$ & $0.102^{\pm.023}$ & $3.095^{\pm.013}$ & $9.315^{\pm.074}$ & $2.727^{\pm.069}$ \\
& 1.00 & $0.464^{\pm.004}$ & $0.650^{\pm.004}$ & $0.740^{\pm.004}$ & $0.132^{\pm.022}$ & $3.090^{\pm.012}$ & $9.345^{\pm.070}$ & $2.575^{\pm.070}$ \\
& 1.25  & $0.467^{\pm.004}$ & $0.653^{\pm.003}$ & $0.743^{\pm.003}$ & $0.101^{\pm.020}$ & $3.088^{\pm.011}$ & $9.355^{\pm.071}$ & $2.576^{\pm.068}$ \\
& 1.50 & $0.466^{\pm.004}$ & $0.654^{\pm.004}$ & $0.745^{\pm.004}$ & $0.173^{\pm.021}$ & $3.090^{\pm.011}$ & $9.350^{\pm.069}$ & $2.573^{\pm.071}$ \\
& 1.75  & $0.462^{\pm.004}$ & $0.648^{\pm.004}$ & $0.737^{\pm.004}$ & $0.152^{\pm.023}$ & $3.092^{\pm.012}$ & $9.338^{\pm.072}$ & $2.571^{\pm.070}$ \\ \midrule

\multirow{2}{*}{\textbf{$\epsilon_1$}}& \multirow{2}{*}{\textbf{$\omega$}} & \multicolumn{3}{c|}{\textbf{R Precision ↑}} & \multirow{2}{*}{\textbf{FID ↓}} & \multirow{2}{*}{\textbf{MM Dist ↓}} & \multirow{2}{*}{\textbf{Diversity →}} & \multirow{2}{*}{\textbf{MM ↑}} \\
& & Top 1 & Top 2 & Top 3 &  &  &  & \\
\midrule

\multirow{5}{*}{\textbf{10\%}} 
& 0.75  & $0.446^{\pm.005}$ & $0.643^{\pm.005}$ & $0.726^{\pm.004}$ & $0.143^{\pm.023}$ & $3.135^{\pm.013}$ & $9.853^{\pm.074}$ & $2.767^{\pm.069}$ \\
& 1.00 & $0.461^{\pm.004}$ & $0.643^{\pm.004}$ & $0.737^{\pm.004}$ & $0.134^{\pm.022}$ & $3.103^{\pm.012}$ & $9.338^{\pm.070}$ & $2.687^{\pm.070}$ \\
& 1.25  & $0.465^{\pm.004}$ & $0.646^{\pm.003}$ & $0.739^{\pm.003}$ & $0.165^{\pm.020}$ & $3.132^{\pm.011}$ & $9.285^{\pm.071}$ & $2.523^{\pm.068}$ \\
& 1.50 & $0.461^{\pm.004}$ & $0.649^{\pm.004}$ & $0.742^{\pm.004}$ & $0.198^{\pm.021}$ & $3.138^{\pm.011}$ & $9.380^{\pm.069}$ & $2.543^{\pm.071}$ \\
& 1.75  & $0.454^{\pm.004}$ & $0.643^{\pm.004}$ & $0.737^{\pm.004}$ & $0.182^{\pm.023}$ & $3.132^{\pm.012}$ & $9.398^{\pm.072}$ & $2.592^{\pm.070}$ \\ \midrule

\multirow{2}{*}{\textbf{CFG}}& \multirow{2}{*}{\textbf{$\omega$}} & \multicolumn{3}{c|}{\textbf{R Precision ↑}} & \multirow{2}{*}{\textbf{FID ↓}} & \multirow{2}{*}{\textbf{MM Dist ↓}} & \multirow{2}{*}{\textbf{Diversity →}} & \multirow{2}{*}{\textbf{MM ↑}} \\
& & Top 1 & Top 2 & Top 3 &  &  &  & \\
\midrule

\multirow{5}{*}{\textbf{}} 
& 6.5  & $0.457^{\pm.004}$ & $0.656^{\pm.004}$ & $0.737^{\pm.004}$ & $0.133^{\pm.023}$ & $3.088^{\pm.012}$ & $9.285^{\pm.072}$ & $2.523^{\pm.070}$ \\
\bottomrule
\end{tabular}
\caption{Parameter Study on $\omega$ and Dropout. $\uparrow$ indicates higher is better, and $\downarrow$ lower is better.}
\label{tab:Hyperparameters_2}
\end{table*}

Across all groups, we systematically sweep the weighting parameter $\omega$ to determine optimal influence magnitudes for each degraded condition as shown in Table~\ref{tab:Hyperparameters_2}. We observe that dropout-only perturbation leads to more stable training compared to noise-based alternatives. This is likely because dropout removes a subset of the conditional inputs while preserving the semantic consistency of the remaining tokens. In contrast, noise injection distorts the content of the condition embeddings, potentially introducing semantic ambiguity and interfering with effective supervision. Moreover, dropout provides a natural curriculum for gradually increasing conditional strength, which is more conducive to stable convergence.

\subsection{Evaluation of Loss Function}
We conduct a comparative study between our proposed DASH Loss and infoNCE loss to evaluate their impact on motion generation quality. While cosine loss encourages alignment between motion and video features at the token level, it lacks explicit structural regularization and fails to preserve the internal relationships within each modality. In contrast, DASH Loss incorporates both token-level similarity and pairwise structural consistency, promoting better semantic grounding and distribution alignment. As shown in Table \ref{tab:comparison_loss}, our method achieves improved performance across all key metrics, demonstrating its effectiveness in bridging the modality gap and enhancing generation quality.
\begin{table*}[htbp]
\centering
\small
\renewcommand{\arraystretch}{1}
\setlength{\tabcolsep}{4.5pt}  
\begin{tabular}{l|ccc|c|c|c|c}
\toprule
\multirow{2}{*}{\textbf{Method}} & \multicolumn{3}{c|}{\textbf{R Precision ↑}} & \multirow{2}{*}{\textbf{FID ↓}} & \multirow{2}{*}{\textbf{MM Dist ↓}} & \multirow{2}{*}{\textbf{Diversity →}} & \multirow{2}{*}{\textbf{MM ↑}} \\
 & Top 1 & Top 2 & Top 3 &  &  &  & \\
\midrule
Real-filtering    & $0.490^{\pm.003}$ & $0.684^{\pm.003}$ & $0.772^{\pm.002}$ & $0.002^{\pm.000}$ & $2.954^{\pm.010}$ & $9.492^{\pm.002}$ & -- \\ \midrule
infoNCE loss       & $0.458^{\pm.003}$ & $0.642^{\pm.003}$ & $0.746^{\pm.003}$ & $1.773^{\pm.012}$ & $3.131^{\pm.010}$ & $9.583^{\pm.071}$ & $2.632^{\pm.071}$ \\ 
Token-wise Margin Loss       & $0.473^{\pm.003}$ & $0.665^{\pm.003}$ & $0.756^{\pm.003}$ & $0.096^{\pm.012}$ & $3.102^{\pm.010}$ & $9.534^{\pm.071}$ & $2.535^{\pm.071}$ \\ \midrule
DASH Loss       & $0.474^{\pm.003}$ & $0.668^{\pm.003}$ & $0.762^{\pm.003}$ & $0.084^{\pm.012}$ & $3.089^{\pm.010}$ & $9.527^{\pm.071}$ & $2.576^{\pm.071}$ \\ 

\bottomrule
\end{tabular}
\caption{ 
Evaluation of Loss Function. $\uparrow$ indicates higher is better, $\downarrow$ indicates lower is better, and $\rightarrow$ indicates closer is better. 
}
\label{tab:comparison_loss}
\end{table*}

\subsection{Supplementary Data on Multimodal Fusion Strategies}
\label{Supplementary Data on Multimodal Fusion Strategies}
We provide the complete results of the ablation studies on multimodal fusion strategies for reference, see Table \ref{tab:fusion_strategies_full}. These supplementary results offer a more comprehensive understanding of how different fusion methods perform under various conditions, and further support the analysis of the sources contributing to performance improvements.

\begin{table*}[htbp]
\centering
\footnotesize
\renewcommand{\arraystretch}{0.8}
\begin{tabular}{l|ccc|c|c|c|c}
\toprule
\multirow{2}{*}{\textbf{Method}} & \multicolumn{3}{c|}{\textbf{R Precision ↑}} & \multirow{2}{*}{\textbf{FID ↓}} & \multirow{2}{*}{\textbf{MM Dist ↓}} & \multirow{2}{*}{\textbf{Diversity →}} & \multirow{2}{*}{\textbf{MM ↑}} \\
 & Top 1 & Top 2 & Top 3 &  &  &  & \\
\midrule
Real-filtering    & $0.490^{\pm.003}$ & $0.684^{\pm.003}$ & $0.772^{\pm.002}$ & $0.002^{\pm.000}$ & $2.954^{\pm.010}$ & $9.492^{\pm.002}$ & -- \\ \midrule
Concat      & $0.463^{\pm.003}$ & $0.652^{\pm.003}$ & $0.742^{\pm.003}$ & $0.192^{\pm.012}$ & $3.296^{\pm.010}$ & $9.687^{\pm.071}$ & $2.412^{\pm.071}$ \\
\hspace*{1em} + Cross-Attn   & $0.380^{\pm.002}$ & $0.568^{\pm.002}$ & $0.684^{\pm.002}$ & $0.707^{\pm.024}$ & $3.652^{\pm.010}$ & $9.308^{\pm.071}$ & \bm{$3.276^{\pm.018}$} \\ \midrule
Concat       & $0.463^{\pm.003}$ & $0.652^{\pm.003}$ & $0.742^{\pm.003}$ & $0.192^{\pm.012}$ & $3.296^{\pm.010}$ & $9.687^{\pm.071}$ & $2.412^{\pm.071}$ \\
\hspace*{1em} + FFT         & $0.430^{\pm.003}$ & $0.626^{\pm.003}$ & $0.726^{\pm.003}$ & $0.364^{\pm.012}$ & $3.392^{\pm.010}$ & $9.703^{\pm.071}$ & $2.640^{\pm.071}$ \\ \midrule
Concat       & $0.463^{\pm.003}$ & $0.652^{\pm.003}$ & $0.742^{\pm.003}$ & $0.192^{\pm.012}$ & $3.296^{\pm.010}$ & $9.687^{\pm.071}$ & $2.412^{\pm.071}$ \\
\hspace*{1em} + DMM   & $0.466^{\pm.002}$ & $0.651^{\pm.002}$ & $0.749^{\pm.002}$ & $0.131^{\pm.024}$ & $3.132^{\pm.010}$ & $9.643^{\pm.071}$ & $2.346^{\pm.018}$ \\ \midrule
Concat      & $0.463^{\pm.003}$ & $0.652^{\pm.003}$ & $0.742^{\pm.003}$ & $0.192^{\pm.012}$ & $3.296^{\pm.010}$ & $9.687^{\pm.071}$ & $2.412^{\pm.071}$ \\
\hspace*{1em} + Self-Attn    & $0.433^{\pm.003}$ & $0.622^{\pm.003}$ & $0.714^{\pm.003}$ & $0.222^{\pm.012}$ & $3.314^{\pm.010}$ & $9.763^{\pm.071}$ & $2.380^{\pm.071}$ \\ 
\hspace*{2em} + DMM      & $0.432^{\pm.003}$ & $0.613^{\pm.003}$ & $0.721^{\pm.003}$ & $0.228^{\pm.012}$ & $3.320^{\pm.010}$ & $9.737^{\pm.071}$ & $2.390^{\pm.071}$ \\ \midrule
Hadamard Product           & $0.441^{\pm.003}$ & $0.636^{\pm.003}$ & $0.741^{\pm.003}$ & $0.243^{\pm.012}$ & $3.219^{\pm.010}$ & $9.393^{\pm.071}$ & $2.370^{\pm.071}$ \\
\hspace*{1em} + FFT         & $0.443^{\pm.003}$ & $0.661^{\pm.003}$ & $0.743^{\pm.003}$ & $0.292^{\pm.012}$ & $3.226^{\pm.010}$ & $9.319^{\pm.071}$ & $2.434^{\pm.071}$ \\
\hspace*{2em} + DMM      & $0.438^{\pm.003}$ & $0.623^{\pm.003}$ & $0.727^{\pm.003}$ & $0.280^{\pm.012}$ & $3.311^{\pm.010}$ & $9.901^{\pm.071}$ & $2.347^{\pm.071}$ \\  \midrule
Element-Wise Addition       & $0.452^{\pm.003}$ & $0.632^{\pm.003}$ & $0.747^{\pm.003}$ & $0.168^{\pm.012}$ & $3.388^{\pm.010}$ & $9.617^{\pm.017}$ & $2.321^{\pm.071}$ \\
\hspace*{1em} + FFT         & $0.435^{\pm.003}$ & $0.634^{\pm.003}$ & $0.743^{\pm.003}$ & $0.204^{\pm.012}$ & $3.256^{\pm.010}$ & $9.430^{\pm.017}$ & $2.352^{\pm.071}$ \\ \midrule
Element-Wise Addition       & $0.452^{\pm.003}$ & $0.632^{\pm.003}$ & $0.747^{\pm.003}$ & $0.168^{\pm.012}$ & $3.388^{\pm.010}$ & $9.617^{\pm.017}$ & $2.321^{\pm.071}$ \\
\hspace*{1em}+ DMM         & $0.451^{\pm.003}$ & $0.643^{\pm.003}$ & $0.750^{\pm.003}$ & $0.204^{\pm.012}$ & $3.256^{\pm.010}$ & $9.445^{\pm.017}$ & $2.402^{\pm.071}$ \\ 
\hspace*{2em} + FFT & $0.435^{\pm.003}$ & $0.630^{\pm.003}$ & $0.744^{\pm.003}$ & $0.254^{\pm.012}$ & $3.299^{\pm.010}$ & $9.624^{\pm.017}$ & $2.402^{\pm.071}$ \\ \midrule
Element-Wise Addition       & $0.452^{\pm.003}$ & $0.632^{\pm.003}$ & $0.747^{\pm.003}$ & $0.168^{\pm.012}$ & $3.388^{\pm.010}$ & $9.617^{\pm.017}$ & $2.321^{\pm.071}$ \\
\hspace*{1em}+ DMM         & $0.451^{\pm.003}$ & $0.643^{\pm.003}$ & $0.750^{\pm.003}$ & $0.204^{\pm.012}$ & $3.256^{\pm.010}$ & $9.445^{\pm.017}$ & $2.402^{\pm.071}$ \\ 
\hspace*{1em}+ FFT & $0.453^{\pm.003}$ & $0.648^{\pm.003}$ & $0.750^{\pm.003}$ & $0.163^{\pm.012}$ & $3.178^{\pm.010}$ & $9.691^{\pm.071}$ & $2.447^{\pm.071}$ \\ 
\hspace*{1em}+ Identity     & $0.459^{\pm.003}$ & $0.652^{\pm.003}$ & $0.752^{\pm.003}$ & $0.147^{\pm.012}$ & $3.124^{\pm.010}$ & $9.677^{\pm.071}$ & $2.347^{\pm.071}$ \\ 
\hspace*{1em}+ Conv (DUET)  &\bm{$0.473^{\pm.003}$}&\bm{$0.664^{\pm.003}$} &\bm{$0.755^{\pm.003}$} & \bm{$0.101^{\pm.024}$} &\bm{$3.087^{\pm.010}$} & \bm{$9.472^{\pm.071}$} & $2.460^{\pm.071}$ \\
\bottomrule
\end{tabular}
\caption{Performance comparison of different multimodal fusion strategies. Table indentation denotes the sequential integration of modules, with each indented block representing a component appended downstream within the overall architecture. The top results in each column are highlighted with 
\textbf{bold} (best).}
\label{tab:fusion_strategies_full}
\end{table*}

\subsection{Quantitative Evaluation of the Video Encoders}
\label{Quantitative Evaluation of the Video Encoders}
To gain deeper insights into the effectiveness and robustness of our framework, we conduct a set of ablation studies aimed at understanding the impact of fine-tuning and model scale on motion generation quality, see Table \ref{tab:video_encoder}. These factors are critical for evaluating the model's generalization ability and its applicability under different resource constraints.

We begin by examining the role of fine-tuning. Specifically, we use the VideoMAEv2-based ViT-G model to perform motion inference directly, without applying any fine-tuning on the virtual skinned motion video dataset. This setup allows us to assess the model’s zero-shot performance and its inherent capacity to generalize. Following this, we study the influence of model size by fine-tuning a smaller ViT-B model that has been distilled from the ViT-G variant, using the same training configuration. This comparison enables us to evaluate the trade-offs between model capacity, computational efficiency, and motion generation quality, providing valuable insights for selecting suitable architectures in practical scenarios.

\begin{table*}[htbp]
\centering
\small
\renewcommand{\arraystretch}{1}
\begin{tabular}{l|ccc|c|c|c|c}
\toprule
\multirow{2}{*}{\textbf{Method}} & \multicolumn{3}{c|}{\textbf{R Precision ↑}} & \multirow{2}{*}{\textbf{FID ↓}} & \multirow{2}{*}{\textbf{MM Dist ↓}} & \multirow{2}{*}{\textbf{Diversity →}} & \multirow{2}{*}{\textbf{MM ↑}} \\
 & Top 1 & Top 2 & Top 3 &  &  &  & \\
\midrule
Real             & $0.511^{\pm.003}$ & $0.703^{\pm.003}$ & $0.797^{\pm.003}$ & $0.002^{\pm.000}$ & $2.974^{\pm.008}$ & $9.503^{\pm.000}$ & --- \\ \midrule
MLD (Baseline)   & $0.481^{\pm.003}$ & $0.673^{\pm.003}$ & $0.772^{\pm.002}$ & $0.473^{\pm.013}$ & $3.196^{\pm.010}$ & $9.724^{\pm.082}$ & $2.413^{\pm.079}$ \\ 
VIT-G with fine-turing  & $0.497^{\pm.003}$ & $0.698^{\pm.003}$  & $0.795^{\pm.003}$ &  $0.179^{\pm 0.024}$  & $3.154^{\pm.010}$  & $9.532^{\pm.080}$ &  $2.496^{\pm.018}$  \\
VIT-G without fine-turing    & $0.446^{\pm.003}$ & $0.643^{\pm.003}$ & $0.751^{\pm.003}$ & $0.238^{\pm.024}$ & $3.334^{\pm.010}$ & $9.653^{\pm.071}$ & $2.654^{\pm.018}$ \\ 
VIT-B with fine-turing    & $0.486^{\pm.003}$ & $0.679^{\pm.003}$ & $0.782^{\pm.003}$ & $0.182^{\pm.012}$ & $3.178^{\pm.010}$ & $9.574^{\pm.071}$ & $2.438^{\pm.071}$ \\

\bottomrule
\end{tabular}
\caption{Performance Assessment of the Video Encoders.
$\uparrow$ indicates higher is better, $\downarrow$ indicates lower is better, and $\rightarrow$ indicates closer is better.}
\label{tab:video_encoder}
\end{table*}

\subsection{Evaluation on Each Component}
\label{Evaluation on each component}
In this section, we provide additional quantitative results and analyses to complement those presented in the main text, see Table \ref{tab:ablation_D.6.}. These supplementary results facilitate a more comprehensive evaluation of each component in our framework.
\begin{table*}[h]
\centering
\small
\renewcommand{\arraystretch}{1}
\begin{tabular}{l|ccc|c|c|c|c}
\toprule
\multirow{2}{*}{\textbf{}} & \multicolumn{3}{c|}{\textbf{R Precision ↑}} & \multirow{2}{*}{\textbf{FID ↓}} & \multirow{2}{*}{\textbf{MM Dist ↓}} & \multirow{2}{*}{\textbf{Diversity →}} & \multirow{2}{*}{\textbf{MM ↑}} \\
 & Top 1 & Top 2 & Top 3 &  &  &  & \\
\midrule
Real       & $0.511^{\pm.003}$ & $0.703^{\pm.003}$ & $0.797^{\pm.003}$ & $0.002^{\pm.000}$ & $2.974^{\pm.008}$ & $9.503^{\pm.065}$ & -- \\ \midrule
Baseline           & $0.481^{\pm.003}$ & $0.673^{\pm.003}$ & $0.772^{\pm.002}$ & $0.473^{\pm.024}$ & $3.196^{\pm.010}$ & $9.724^{\pm.071}$ & $2.413^{\pm.018}$ \\
+ Filtering    & $0.446^{\pm.003}$ & $0.628^{\pm.003}$ & $0.734^{\pm.002}$ & $0.396^{\pm.024}$ & $3.156^{\pm.010}$ & $9.710^{\pm.071}$ & $2.433^{\pm.018}$ \\ \midrule
Real-filtering    & $0.490^{\pm.003}$ & $0.684^{\pm.003}$ & $0.772^{\pm.002}$ & $0.002^{\pm.000}$ & $2.954^{\pm.010}$ & $9.492^{\pm.081}$ & -- \\ \midrule
+ Video            & $0.463^{\pm.003}$ & $0.652^{\pm.003}$ & $0.742^{\pm.003}$ & $0.192^{\pm.012}$ & $3.296^{\pm.010}$ & $9.687^{\pm.071}$ & $2.412^{\pm.071}$ \\
+ DUET   & $0.473^{\pm.003}$ & $0.664^{\pm.003}$ & $0.755^{\pm.003}$ & $0.101^{\pm.024}$ &\bm{$3.087^{\pm.010}$}  & $9.472^{\pm.071}$ & $2.460^{\pm.071}$ \\
+ DASH Loss  & \bm{$0.474^{\pm.003}$} & \bm{$0.668^{\pm.003}$} &\bm{$0.764^{\pm.003}$}  & \bm{$0.084^{\pm.012}$} & $3.089^{\pm.010}$ & \bm{$9.527^{\pm.071}$}& \bm{$2.576^{\pm.071}$} \\
\bottomrule
\end{tabular}
\caption{Evaluation on each component. The top results are highlighted in each column with 
\textbf{bold}.}
\label{tab:ablation_D.6.}
\end{table*}

\subsection{Inference Time} 
The model’s inference statistics indicate approximately 7960 GFLOPs and 7932 GMACs per forward pass, representing the total number of floating-point and multiply-accumulate operations required to process each input. All inference experiments were conducted on a single NVIDIA A100 GPU with 80GB of memory. Under this configuration, the average inference time per sample (AITS) was observed to range from approximately 0.092 seconds, reflecting efficient runtime performance and effective hardware utilization, particularly in batch processing scenarios.

\subsection{Model Parameter Statistics}
To provide a comprehensive overview of our model architecture, we summarize the major components and their corresponding parameter counts in Table \ref{tab:params}. The entire system consists of multiple encoders and decoders tailored for vision, text, and motion modalities. Notably, the largest component is the pretrained Vision Transformer encoder, containing 953M parameters, which remains frozen during training. Among all modules, only 21.6M parameters are trainable, ensuring efficient optimization while leveraging powerful pretrained backbones.

\begin{table}[htbp]
\centering
\begin{tabular}{l l c}
\toprule
\textbf{Module Name} & \textbf{Component} & \textbf{Param. Count} \\
\midrule
pretrainVisionTransformerEncoder & VisionTransformer & 953 M \\
text\_encoder & MldTextEncoder & 427 M \\
vae & MldVae & 18.8 M \\
denoiser & MldDenoiser & 21.6 M \\
t2m\_textencoder & TextEncoderBiGRUCo & 4.1 M \\
t2m\_moveencoder & MovementConvEncoder & 1.8 M \\
t2m\_motionencoder & MotionEncoderBiGRUCo & 15.7 M \\
\bottomrule
\end{tabular}
\caption{Model components and parameter statistics. "Trainable params" refer to parameters updated during training.}
\label{tab:params}
\end{table}

Overall, by freezing the majority of the parameters (1.4B non-trainable) and optimizing only a lightweight subset (21.6M trainable), our method strikes a balance between parameter efficiency and representation power.
\section{Video Motion Dataset}
\label{Video Motion Dataset}
\subsection{Overview of the HumanML3D Dataset}
The HumanML3D dataset provides a comprehensive and standardized representation of human motion, focusing on skeleton-level analysis. Each motion sequence is stored as a NumPy array with 263-dimensional features per frame, capturing both rotation-invariant and rotation-related information, including joint positions, velocities, angular changes, and joint rotations. Instead of using raw Skinned Multi-Person Linear (SMPL) parameters, the dataset represents motion through a consistent 22-joint skeleton structure with normalized body shape across all samples. By intentionally excluding skinned mesh data, textures, and clothing, HumanML3D emphasizes clean skeletal motion suitable for tasks involving motion understanding rather than detailed 3D surface rendering.

\vspace{5pt}\noindent\textbf{Motion Data Representation.}
The HumanML3D dataset offers an extensive repository of motion data seamlessly integrated with vivid natural language descriptions, stored as NumPy arrays and text files. Each motion sequence is elegantly organized as an M×263 matrix, where M signifies the number of frames. Every 263-dimensional feature vector per frame encapsulates a sophisticated array of rotation-invariant and rotation-related attributes, including root joint angular velocity, translation velocity, vertical displacement, local joint positions and velocities, 6D joint rotation representations, and binary foot contact indicators, forming a robust foundation for advanced motion analysis.

\vspace{5pt}\noindent\textbf{Standardized Joint-Based Motion Features.}
Uniquely, HumanML3D refrains from including raw SMPL parameters such as pose, shape, or translation. Instead, it transforms motion data into standardized joint sequences and derived features, utilizing the 22-joint structure of the SMPL skeleton to precisely articulate human poses. Each frame is defined by accurate 3D coordinates for these joints. Shape parameters are deliberately uniform, with all motion data normalized to a consistent human template, ensuring no variations in body shape across samples for streamlined analysis.

\vspace{5pt}\noindent\textbf{Skeleton-Level Data Without Skinning.}
HumanML3D is intentionally crafted to focus exclusively on skeleton-level motion data, explicitly excluding skinned 3D body mesh models or skinning processes. It omits mesh vertex sequences, FBX files, texture maps, and clothing models, prioritizing skeletal motion data and its associated feature representations over skinned vertex clouds or fully animated mesh sequences. This deliberate exclusion of skinning underscores HumanML3D’s suitability for applications centered on skeletal motion analysis rather than detailed 3D mesh rendering or skinning-dependent visualizations.
\begin{figure}[htbp]
    \centering
    \includegraphics[width=0.7\linewidth]{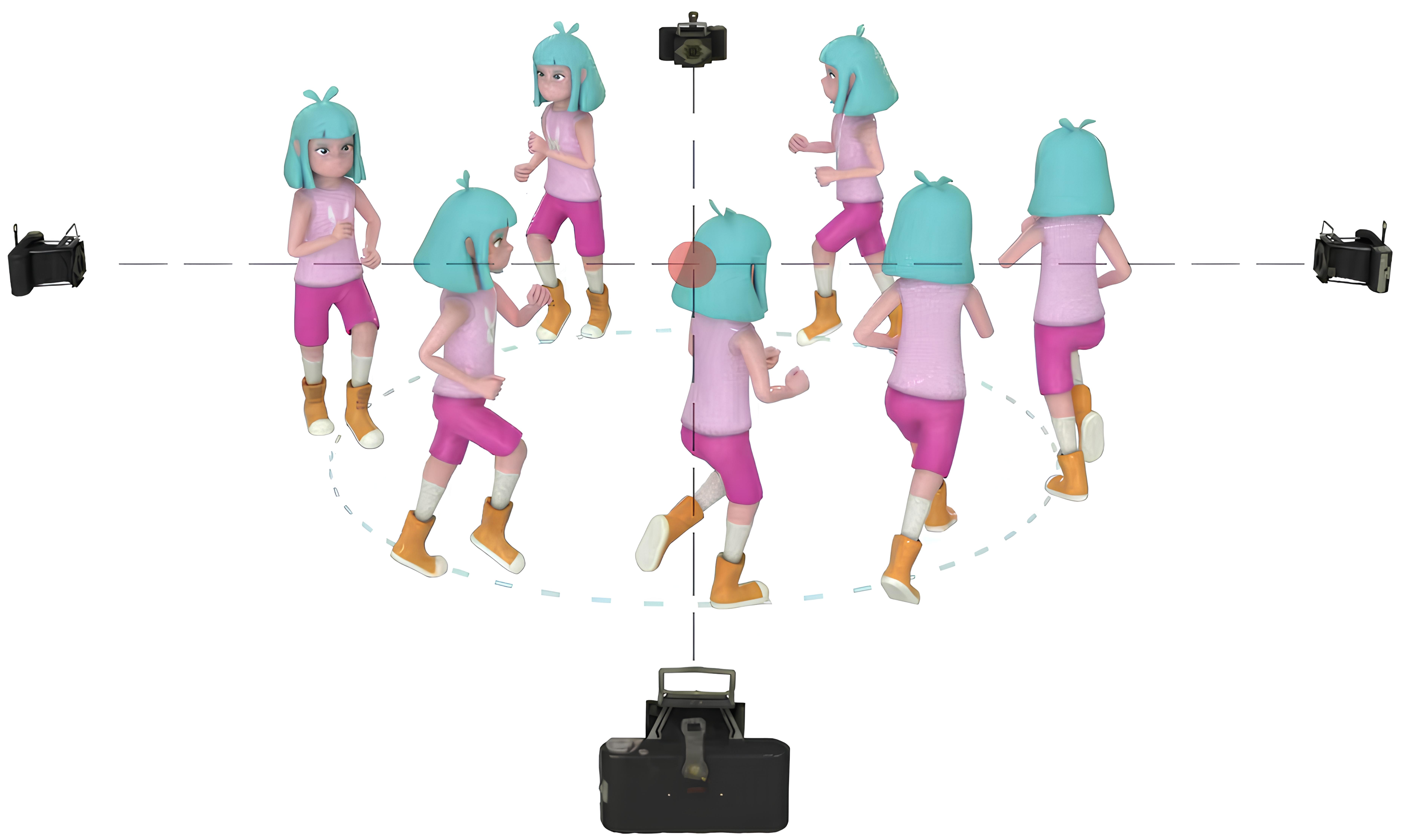}
    \caption{Video motion dataset creation workflow visualization.}
    \label{fig:Video motion dataset creation workflow visualization.}
\end{figure}
\subsection{HumanML3D Visualization}
To achieve visualization and in-depth analysis of the HumanML3D dataset, we first converted the .npy files into Biovision Hierarchy (BVH) format for convenient visualization using Blender software. However, skeletal-based virtual human motions often lack realism. Therefore, we further converted the BVH-format data into SMPL model format and applied skinning to enhance the visual authenticity of the motion.

Notably, the BVH format utilizes a skeletal structure comprising 17 joints, whereas the SMPL model includes 22 joints. The five additional joints in the SMPL model correspond to vertices at the extremities of the limbs and the top of the head. During conversion, to ensure compatibility, the values for these additional joints were set to zero. After generating the SMPL models, we assigned skinning weights and standardized the initial human shape to an A-pose to maintain consistency and standardization.

Subsequently, we converted the SMPL-format data into FBX format and utilized Blender software to set up four virtual cameras, capturing the motion sequences comprehensively from the east, south, west, and north directions, see Fig. \ref{fig:Video motion dataset creation workflow visualization.}. This process yielded a total of 116,800 video motion videos. To ensure high data quality, we employed the data cleaning approach described in Appendix B, ultimately obtaining 71,220 video motion videos with limited but inevitable errors, accounting for approximately 61\% of the original dataset. The entire process took 45 days to complete, utilizing four NVIDIA RTX 4090 GPUs to ensure efficient and high-fidelity rendering and processing.

\subsection{Anomaly Data Analysis}
\label{Anomaly Data Analysis.}
\begin{figure}[htbp]
    \centering
    \includegraphics[width=0.8\textwidth]{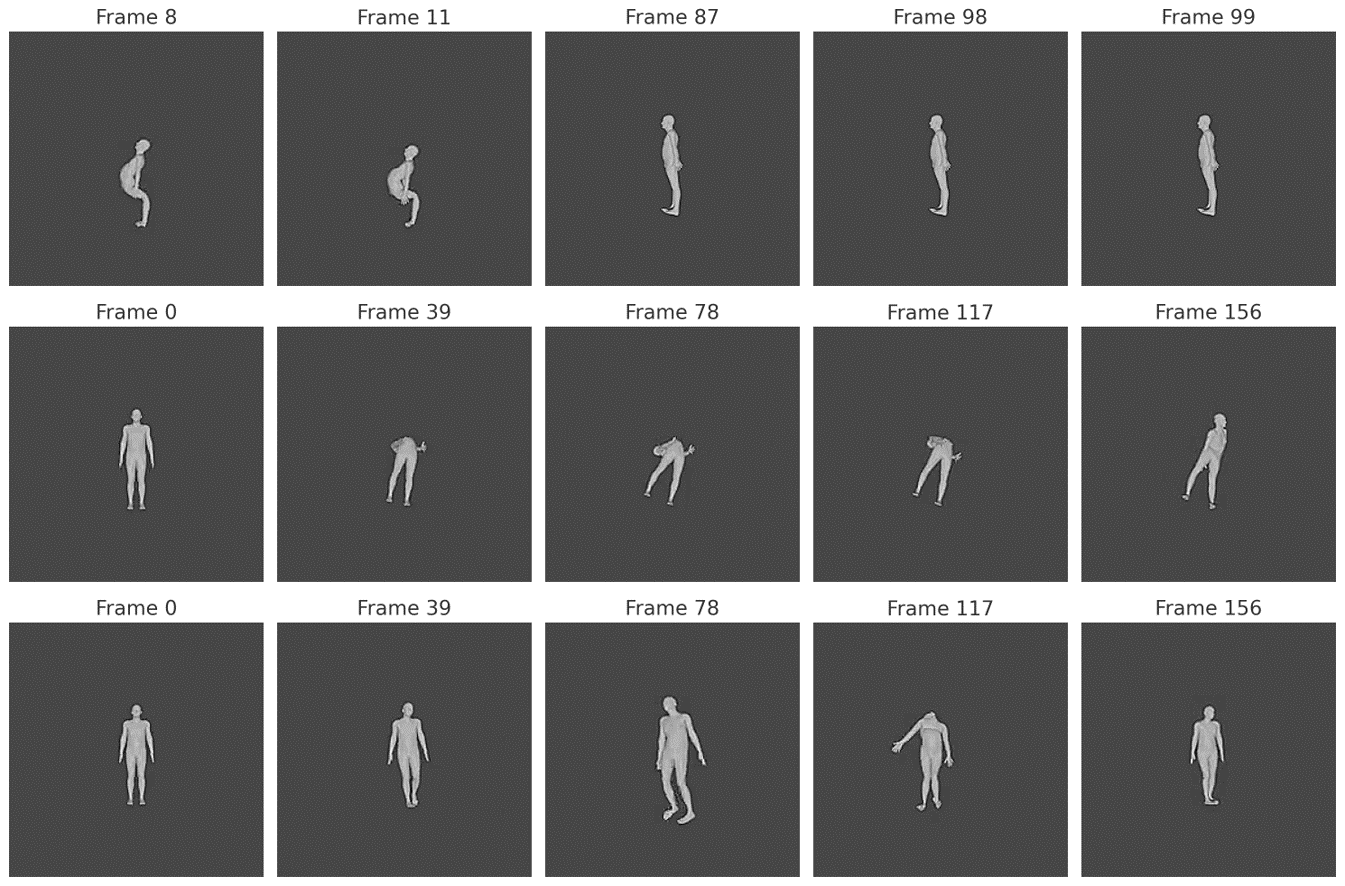}
    \caption{Skinning errors.}
    \label{fig:img1}
\end{figure}

\begin{figure}[htbp]
    \centering
    \includegraphics[width=0.8\textwidth]{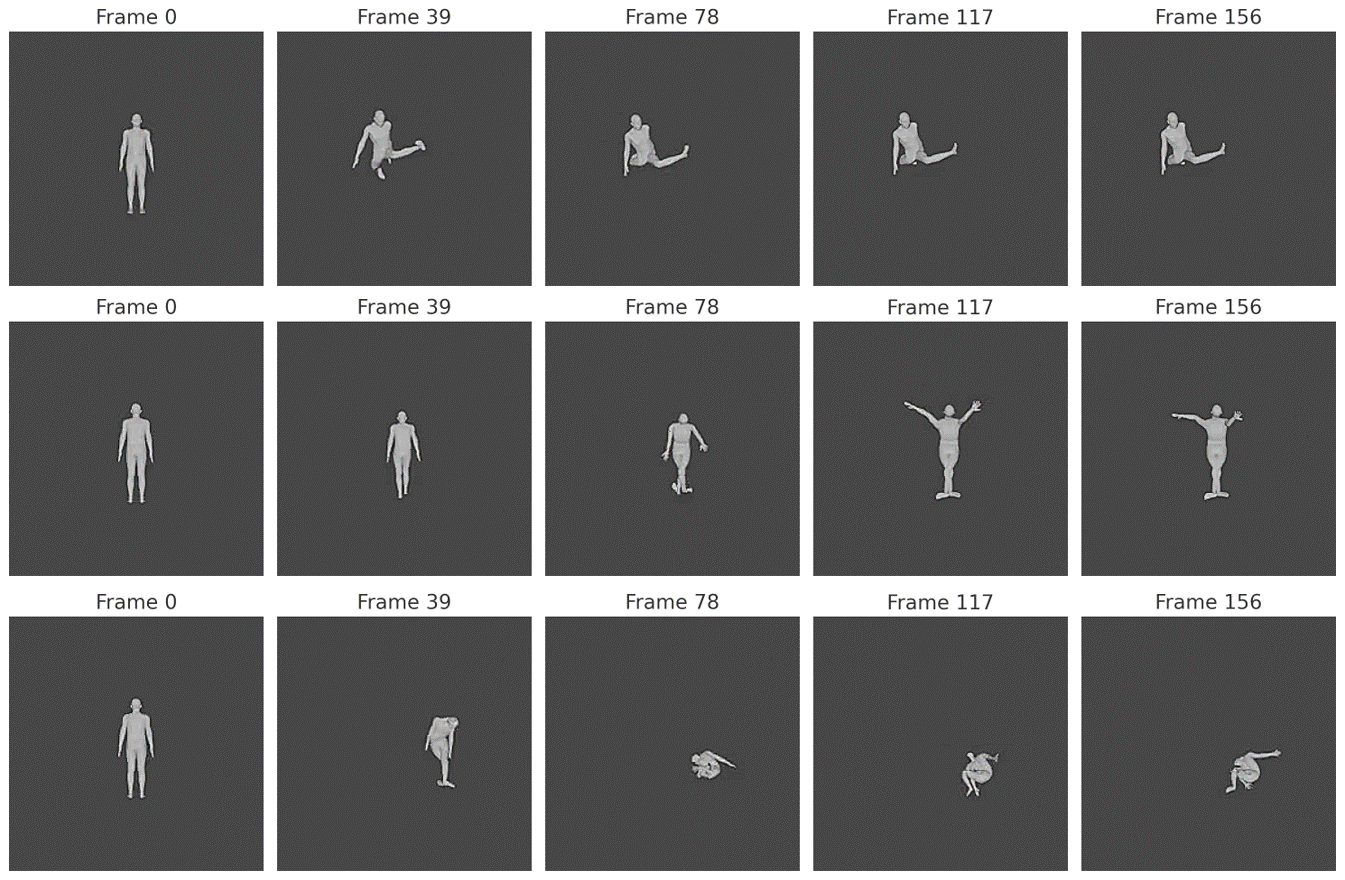}
    \caption{Data quality issues.}
    \label{fig:img2}
\end{figure}

\begin{figure}[htbp]
    \centering
    \includegraphics[width=0.8\textwidth]{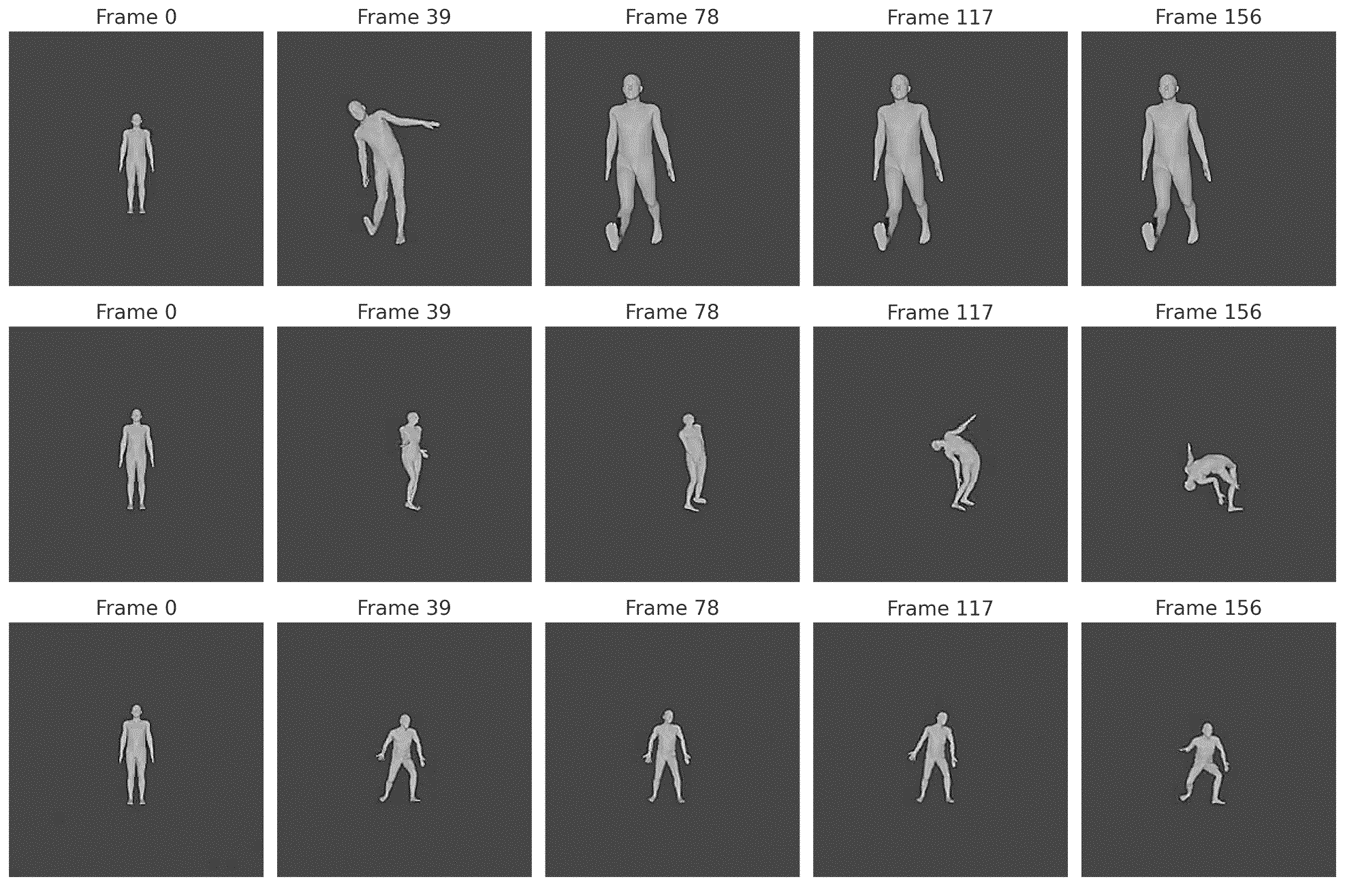}
    \caption{Mild deviations in the motion itself.}
    \label{fig:img3}
\end{figure}

Based on the analysis results, anomalous motion samples account for 39\% of the entire dataset. 
These refer to video clips that, after automatic preprocessing by our data-cleaning script, still contain artifacts or motion inconsistencies, and are thus categorized as anomalous motion samples. Through visualization analysis, we were able to identify these anomalous samples and began investigating the reasons behind such a high anomaly rate. We categorized the anomalous data into three main types:
\begin{itemize}
    \item \textbf{Skinning errors}, which result in incorrect or inverted skin deformations of the human body, as shown in Fig. \ref{fig:img1};
    \item \textbf{Data quality issues}, where the overall motion appears generally normal but contains locally unbalanced or disproportionate movements, as shown in Fig. \ref{fig:img2};
    \item \textbf{Mild deviations in the motion itself}, where the motion sequence displays subtle but noticeable unnatural or unrealistic elements, as shown in Fig. \ref{fig:img3}.

\end{itemize}
In the future, we plan to further improve our visualization methods by integrating more advanced techniques to gain a deeper understanding of and better monitor the quality of motion generation. These enhancements will help us identify deficiencies in the data creation process and guide the refinement of both generation and curation pipelines. Ultimately, this will facilitate the production of higher-quality motion samples.

\end{document}